\documentclass[a4paper,12pt]{article}
\usepackage{latexsym}
\usepackage{comment}
\usepackage[english]{babel}
\usepackage[font={small}]{caption}
\usepackage{epsfig,amssymb,euscript,slashed}
\usepackage[linktocpage]{hyperref}
\usepackage{amsmath}
\usepackage[noadjust]{cite}
\usepackage{mathtools}
\usepackage{mathrsfs}
\usepackage{physics}
\usepackage{extarrows}
\usepackage{xcolor}
\usepackage{float}
\usepackage{graphicx}
\usepackage{caption}
\usepackage{subcaption}
\usepackage{tcolorbox}
\usepackage{tikz}
\usetikzlibrary{positioning,shapes.multipart,arrows.meta,decorations.pathreplacing,calligraphy,patterns}
\usepackage{adjustbox}
\usepackage{nicematrix}
\usepackage{array,calc}
\usepackage[]{graphicx}
\usepackage{color}
\usepackage[final]{showkeys} 
\usepackage{bbm}
\usepackage{fancybox}
\usepackage{ytableau}
\usepackage{enumerate}
\usepackage[shortlabels]{enumitem}
\usepackage{IEEEtrantools}

\numberwithin{equation}{section}

\def\Ad{\dot{A}}
\def\Bd{\dot{B}}
\def\Cd{\dot{C}}
\def\Dd{\dot{D}}

\def\mt{\tilde{m}}

\def\ep{\epsilon}

\def\ald{\dot{\alpha}}

\def\yt{\tilde{y}}

\def\lam{\lambda}

\def\pd{\partial}

\def\CYon{\ytableausetup{centertableaux}}

\def\YTnormalsize{\ytableausetup{boxsize=1.25ex}}
\def\YTscriptsize{\ytableausetup{boxsize=0.65ex}}

\def\Phibf{\boldsymbol{\Phi}}

\definecolor{dfill1}{rgb}{0.0, 0.45, 0.73}
\definecolor{dfill2}{rgb}{0.0, 0.19, 0.33}
\definecolor{dfill3}{rgb}{0.1, 0.1, 0.5}

\RenewDocumentCommand\ket{s m}{%
   \IfBooleanTF{#1}
     {\left\lvert\smash{#2}\right\rangle}%
     {\left\lvert{#2}\right\rangle}%
}

\newcommand{\bigket}[1]{\bigl|#1\bigr\rangle}

\definecolor{garrow1}{rgb}{0.90,0.75,0.5}

\def\alphad{{\dot{\alpha}}}

\def\Vt{{\tilde{V}}}
\def\half{\frac12}

\def\etad{\dot{\eta}}

\makeatletter
\g@addto@macro\bfseries{\boldmath}
\makeatother

\def\cA{{\cal A}}

\def\cE{{\cal E}}

\def\cG{{\cal G}}

\def\cN{{\cal N}}
\def\cO{{\cal O}}

\def\bbZ{{\mathbb{Z}}}

\usepackage{pict2e}

\makeatletter
\DeclareRobustCommand{\loplus}{\mathbin{\mathpalette\dog@lsemi{+}}}
\DeclareRobustCommand{\lotimes}{\mathbin{\mathpalette\dog@lsemi{\times}}}
\DeclareRobustCommand{\roplus}{\mathbin{\mathpalette\dog@rsemi{+}}}
\DeclareRobustCommand{\rotimes}{\mathbin{\mathpalette\dog@rsemi{\times}}}

\newcommand{\dog@rsemi}[2]{\dog@semi{#1}{#2}{-90,90}}
\newcommand{\dog@lsemi}[2]{\dog@semi{#1}{#2}{270,90}}
\newcommand{\dog@semi}[3]{%
  \begingroup
  \sbox\z@{$\m@th#1#2$}%
  \setlength{\unitlength}{\dimexpr\ht\z@+\dp\z@\relax}%
  \makebox[\wd\z@]{\raisebox{-\dp\z@}{%
    \begin{picture}(1,1)
    \linethickness{\variable@rule{#1}}
    \roundcap
    \put(0.5,0.5){\makebox(0,0){\raisebox{\dp\z@}{$\m@th#1#2$}}}
    \put(0.5,0.5){\arc[#3]{0.5}}
    \end{picture}%
  }}%
  \endgroup
}
\newcommand{\variable@rule}[1]{%
  \fontdimen8  
  \ifx#1\displaystyle\textfont3\else
    \ifx#1\textstyle\textfont3\else
      \ifx#1\scriptstyle\scriptfont3\else
        \scriptscriptfont3\relax
  \fi\fi\fi
}
\makeatother

\begin{document}
\font\cmss=cmss10 \font\cmsss=cmss10 at 7pt

% format
 \baselineskip=18pt  % a la harvmac
 \numberwithin{equation}{section}  % make eq labels (sec.num)

\vspace*{-2cm}
\begin{flushright}{  
\scriptsize YITP-26-43}
\end{flushright}
\hfill
\vspace{18pt}
\begin{center}
{\Large 
\textbf{
Towering Gravitons in AdS$_3$/CFT$_2$}}\\ 
\end{center}

\vspace{8pt}
\begin{center}
{\textsl{
Marcel R. R. Hughes$^{\,a}$,
Kohei Jin$^{\,a}$,
Daiki Matsumoto$^{\,a}$,\\
Leon Miyahara$^{\,a}$
 and Masaki Shigemori$^{\,a,b}$}}

\vspace{1cm}

\textit{\small ${}^a$ Department of Physics, Nagoya University\\
Furo-cho, Chikusa-ku, Nagoya 464-8602, Japan} \\ \vspace{6pt}

\textit{\small ${}^b$ 
Center for Gravitational Physics,\\
Yukawa Institute for Theoretical Physics, Kyoto University\\
Kitashirakawa Oiwakecho, Sakyo-ku, Kyoto 606-8502, Japan
}\\
\vspace{6pt}

\end{center}

\vspace{12pt}

\begin{center}
\textbf{Abstract}
\end{center}

\vspace{4pt} {\small
\noindent
BPS states in holographic CFTs are usually classified into supergravitons, namely BPS fluctuations around empty AdS, and black-hole microstates, which appear above an energy threshold.   In AdS$_3$/CFT$_2$, however, this picture is incomplete because of additional degrees of freedom, called singletons, associated with boundary diffeomorphisms.  
We present a general procedure for extending the BPS spectrum of supergravitons by dressing them with singletons, thereby defining a generalized, gravity-sector Hilbert space that admits decomposition into affine multiplets of the full superconformal algebra.  This extends the procedure previously proposed in \cite{Hughes:2025tdy} which was applicable only at low levels, by removing that limitation.  We apply the new procedure to the D1-D5 CFT ${\rm Sym}^N(T^4)$ and explicitly construct affine multiplets in the gravity sector for the $N=2$ theory up to level $h=2$.
We find that, at the free orbifold point, the gravity-sector spectrum agrees with the CFT up to $h=\frac12$.  Upon turning on a deformation, however, states at $h=1$ lift and the agreement improves to $h=\frac32$.  Interestingly, the lifting occurs between states in the gravity sector, involving mixtures of supergravitons and singletons, and stringy states.  We conjecture that, upon deformation, the gravity-sector Hilbert space becomes the monotone Hilbert space while its complement becomes the fortuitous Hilbert space.
}

\vspace{1cm}

\thispagestyle{empty}

\vfill
\vskip 5.mm
\hrule width 5.cm
\vskip 2.mm
{
\noindent  {\scriptsize e-mails:  {\texttt{\{hughes.mrr, leon, D.Matsumoto, jinkohei\}@eken.phys.nagoya-u.ac.jp, masaki.shigemori@nagoya-u.jp} }}}

\setcounter{footnote}{0}
\setcounter{page}{0}

\newpage

\YTnormalsize\CYon

\tableofcontents

% ========== title (paper version, a la harvmac) ends here ==========

%%%%%%%%%%%%%%%%%%%%%%%%%%%%%%%%%%%%%%%%%%%
%%%           TITLE ENDS HERE
%%%%%%%%%%%%%%%%%%%%%%%%%%%%%%%%%%%%%%%%%%%

%\tableofcontents
%\printindex

%%%%%%%%%%%%%%%%%%%%%%%%%%%%%%%%%%%%%%%%%%%
%%%        MAIN TEXT BEGINS HERE
%%%%%%%%%%%%%%%%%%%%%%%%%%%%%%%%%%%%%%%%%%%

\section{Introduction}
\label{sec:intro}

% \begin{itemize}
% \item AdS/CFT.  BPS states.  there are supergraviton and black hole states.
% \item There are also singletons.  Total affine algebra. 
%     \item  (super)graviton sector and the (super)gravity sector as its completion.  Lifting.  Fig \ref{fig:structure_BPS_Hilb_spc}.
% \item Determining $H_0^{gravity}$.  Previous paper \cite{Hughes:2025tdy}: valid up to certain levels.  New procedure.
% \item Explain briefly what we do: All gravity affine towers up to level 2.  How they compare with the CFT towers. include unrefined PF for giving the idea.
% \end{itemize}

One of the central aspects of developing the holographic dictionary is the matching of states between the two sides of the AdS/CFT correspondence~\cite{Maldacena:1997re}. In doing so, BPS states have played a key role; often benefiting from non-renormalisation theorems, their properties can be studied in the weakly coupling regime of the CFT, extrapolated to strong coupling, and compared with the gravitational theory. Supersymmetric indices are typically invaluable tools in this process.

In this matching of BPS states, it is natural to ask how CFT states divide into those dual to perturbative supergravity modes in global AdS, ``supergraviton states'', and those dual to black hole microstates, ``black hole states''. Despite black hole states existing only above a certain central-charge-scaling energy threshold, such a distinction is not sufficient since graviton states also exist at arbitrarily high energies.\footnote{Once the number of supergraviton excitations considered is of order $\sim c$, their backreaction on global AdS is important. For a coherent supergraviton gas, their backreaction is described by smooth horizonless geometries~\cite{Bena:2016ypk}.}

Specializing to the AdS$_3$/CFT$_2$ holographic duality relevant to the D1-D5 system, the correspondence is between type IIB string theory on AdS$_3\times S^3\times M_4$, with $M_4=T^4$ or $K3$, and the so-called D1-D5 CFT\@. This CFT$_2$ has $\cN=(4,4)$ superconformal symmetry and is strongly believed to be described by the conformal manifold that includes the symmetric product orbifold theory $\mathrm{Sym}^{N}(M_4)$, with central charge $c=6N$. The symmetric orbifold represents the zero coupling description of the D1-D5 CFT\@. This holographic duality was the setting for the first matching of the leading-order large-$N$ growth of the number of black hole microstates between the bulk and CFT sides of the duality~\cite{Strominger:1996sh}. Later on, for $M_4=K3$ the agreement of the CFT elliptic genus (a supersymmetric index for this theory) and the ``supergraviton elliptic genus'' was shown~\cite{deBoer:1998us} to hold below the bound in conformal dimension $h\leq\frac{N+1}{4}$. This supergraviton elliptic genus is not, \textit{a priori}, a protected index; instead it is simply a signed trace over the (suitably enumerated) bulk Kaluza-Klein modes~\cite{Maldacena:1998bw}---the contribution of supergravitons to the elliptic genus. The observed matching of these quantities yields a sharp energy bound at which black hole states begin to exist in the CFT's BPS spectrum. Similarly, for $M_4=T^4$ a similar agreement was observed in~\cite{Maldacena:1999bp} using the modified elliptic genus (MEG), the standard supersymmetric index of the D1-D5 CFT for $T^4$; in this case the agreement\footnote{The agreement of the CFT and supergraviton spectra for the $T^4$ theory using the MEG is relatively simple, due to both quantities only receiving contributions from the global vacuum below the black hole bound. A detailed matching was recently demonstrated in~\cite{Hughes:2026qqn} using a generalization of the MEG, the ``resolved elliptic genus'', which will also be used in this paper.} is below the bound $h<\frac{N}{4}$.

\begin{figure}[tb]
\begin{center}
\begin{tikzpicture}[scale=1.2]
\draw[fill=black!5!white] (0,0) ellipse (4 and 2);
\path[fill=red,opacity=0.3] (-1.35,0) ellipse (2.5 and 1.25);
\path[fill=blue,opacity=0.3] (-2.2,0) ellipse (1.5 and 0.75);
\path[pattern=north east lines,pattern color=black]
  (-0.3,-0.7) ellipse (3 and 1);
\node at (2,2.05) {$H_0^{\rm CFT}$};
\node[color=red] at (0.6,1.2) {$H_0^{\rm gravity}$};
\node[color=blue] at (-0.6,0.7) {$H_0^{\rm graviton}$};
\node at (2.5,0.1) {$V^{\rm lifted}$};
\end{tikzpicture}
\caption{\sl A schematic illustration of the structure of the BPS Hilbert space in the free symmetric orbifold CFT\@.
The BPS graviton Hilbert space, $H_0^{\rm graviton}$, shown in blue, is a subspace of the CFT BPS Hilbert space, $H_0^{\rm CFT}$.  Adding singleton excitations extends $H_0^{\rm graviton}$ to the BPS gravity Hilbert space, $H_0^{\rm gravity}$, shown in red.  Upon turning on the deformation, some set of states in these BPS spaces, $V^{\rm lifted}$, indicated by the hatched region, are lifted and disappear.  
\label{fig:structure_BPS_Hilb_spc}
}
\end{center}
\end{figure}
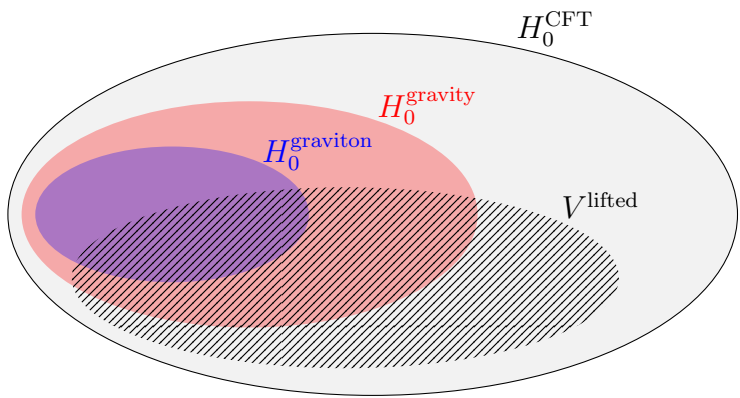

A complication in classifying supergraviton and black hole states in the case of AdS$_3$/CFT$_2$, however, is the existence of so-called singleton states. In the bulk, these represent diffeomorphisms that do not vanish at the AdS boundary~\cite{Brown:1986nw}, while in the CFT they correspond to the affine generators of the superconformal algebra.   In the symmetric orbifold CFT, supergraviton (or simply graviton) states span a subspace $H_0^{\rm graviton}$ of the free-CFT BPS Hilbert space $H_0^{\rm CFT}$ and furnish representations of the global subalgebra of the superconformal algebra, \textit{i.e.}\ global multiplets. Dressing gravitons with singletons, however, extends them to representations of the full superconformal algebra, \textit{i.e.}\ affine multiplets.  We denote the resulting, extended BPS Hilbert space by $H_0^{\rm gravity}$, since it is naturally identified with the space of states realizable in supergravity.  We often refer to $H_0^{\rm graviton}$ and $H_0^{\rm gravity}$ as the \textit{graviton sector} and the \textit{gravity sector}, respectively, and
the affine multiplets in the gravity sector as \textit{gravity towers}.
See Figure~\ref{fig:structure_BPS_Hilb_spc} for a schematic picture of these BPS Hilbert spaces.  The complement of $H_0^{\rm gravity}$ in $H_0^{\rm CFT}$, denoted by $\overline{H_0^{\rm gravity}}$, can be interpreted as the space of BPS black-hole microstates.

The aforementioned matching of supersymmetry indices below the black-hole bound did not fully take singletons into account \cite{deBoer:1998us, Maldacena:1999bp}.  In other words, it was a comparison between supersymmetry indices computed over $H_0^{\rm graviton}$ and those over $H_0^{\rm CFT}$.  It is therefore interesting to ask how the matching of the BPS spectrum is modified once singletons are included.  For that, we need to work out the spectrum of gravity towers in $H_0^{\rm gravity}$.

In this paper, we carry out the program of extending the graviton Hilbert space $H_0^{\rm graviton}$ to the gravity Hilbert space $H_0^{\rm gravity}$ in the $M_4=T^4$ theory for $N=2$ up to level (conformal dimension) $h=2$, by identifying all gravity towers in the gravity sector and writing down the corresponding affine primaries.  This program was initiated in \cite{Hughes:2025tdy}, but due to a limitation in the procedure used there, it was carried out only up to lower levels.\footnote{Up to $h=1$ for $T^4$ and $h=\frac32$ for $K3$.}  In this paper, we propose a new, improved procedure applicable at arbitrary level, and demonstrate its validity by working out explicit gravity towers.

After identifying affine towers in the gravity sector, we compare them with those in the full CFT\@.  At the orbifold point, the Hilbert space of the $T^4$ theory decomposes into symmetry sectors labeled by Young diagrams $\lambda$, as we will review in section \ref{sec:background}.  In the case of $N=2$, where $\lambda=\YTnormalsize\ydiagram{2},\ydiagram{1},\ydiagram{1,1}$, the left-moving part of the BPS partition functions admit the following character expansion:
\begin{subequations} \label{eq:S_CFT_exp}
    \begin{align}
        S^{\mathrm{CFT}}_{\YTscriptsize\ydiagram{2}} &= \mathbf{\Phi}_{0} + 3 \mathbf{\Phi}_{0,1} + \mathbf{\Phi}_{0,2}+ \cdots \ ,\label{eq:S_CFT_2_exp}\\
        S^{\mathrm{CFT}}_{\YTscriptsize\ydiagram{1}} &= \mathbf{\Phi}_{\frac12} - 6 \mathbf{\Phi}_{0,1} - 28\mathbf{\Phi}_{0,2} + \cdots \ ,\label{eq:S_CFT_1_exp}\\
        S^{\mathrm{CFT}}_{\YTscriptsize\ydiagram{1,1}} &= -2 \mathbf{\Phi}_{\frac12} + 8 \mathbf{\Phi}_{0,2} + \cdots \ . \label{eq:S_CFT_1,1_exp}
    \end{align}
\end{subequations}
Here, $\Phibf_j$ and $\Phibf_{j,h}$ are short and long characters of the $\cN=4$ conformal algebra, which correspond to short and long affine primaries, respectively, while $j$ and $h$ are the R-charge and the conformal weight, respectively. The questions is which of these characters can be accounted for within the gravity sector.  We will find that all short characters in \eqref{eq:S_CFT_exp}, as well as the long characters in \eqref{eq:S_CFT_2_exp}, are accounted for within the gravity sector, while none of long characters in 
\eqref{eq:S_CFT_1_exp} or \eqref{eq:S_CFT_1,1_exp} are.  
Thus, the free-theory spectrum of affine multiplets agree between the gravity sector and the full CFT up to $h=\frac12$.

This is not the full story, however, because the above construction of gravity towers was carried out at the free orbifold point and the CFT partition functions \eqref{eq:S_CFT_exp} are likewise defined over the free CFT Hilbert space $H_0^{\rm CFT}$.  Upon deforming away from the orbifold point, some of the BPS states lift and hence disappear from the BPS spectrum; see the hatched region in Figure~\ref{fig:structure_BPS_Hilb_spc}.  In fact, we will find that, once interaction is turned on, the long gravity towers in the $\lambda=\YTnormalsize\ydiagram{2}$ sector, which belong to $H_0^{\rm gravity}$, combine with the stringy states in $\overline{H_0^{\rm gravity}}$ and lift.  If we take this phenomenon into account, we find that the spectrum of affine multiplets agree between the gravity sector and the full CFT up to $h=\frac32$ (although the agreement persists up to $h=2$ in the $\lambda=\ydiagram{2}$ sector).

The recent ideas of ``fortuity'' \cite{Chang:2022mjp, Chang:2024zqi} refines the notion of separating BPS states in the interacting CFT into supergraviton and black hole types. This classification is based on the categories of ``monotone'' states, which remain BPS for all $N$, including $N=\infty$, and ``fortuitous'' states, which are BPS for only a finite range of $N$. These ideas have now been studied in various examples of holography~\cite{Chang:2013fba,Chang:2022mjp,Choi:2022caq, Choi:2023vdm, Chang:2023zqk, Choi:2023znd, Budzik:2023vtr,Chang:2024lxt,Chang:2025rqy,Chang:2025wgo,Kim:2025vup,Behan:2025hbx,Choi:2025pqr,Belin:2025hsg,Chen:2025sum}, both in terms of explicit BPS states and supercharge cohomologies~\cite{Kinney:2005ej, Grant:2008sk, Chang:2013fba}.
It is conjectured that monotone CFT states correspond in the bulk to perturbative supergravity excitations and smooth horizonless geometries (see the reviews~\cite{Bena:2022ldq,Bena:2022rna,Shigemori:2020yuo} and references therein), whereas fortuitous states correspond to black hole microstates. 
In particular, singleton states, being boundary diffeomorphisms,
are classified as monotone~\cite{Chang:2025rqy}, and hence so are all states in gravity towers.
Based on the findings in this paper, we conjecture that the gravity sector of the interacting theory gives precisely the sector of monotone states, while its complement corresponds to fortuitous states.

The organization of the rest of the paper is as follows.
In Section~\ref{sec:background}, we review the aspects of the D1-D5 CFT on $T^4$ relevant to the present paper, including graviton states, singleton states, and the Schur-Weyl decomposition of the Hilbert space and the partition functions.
In Section~\ref{sec:method}, we present a general procedure to find all gravity towers, valid at arbitrary level.
In Section~\ref{sec:results}, we apply the general procedure to the $T^4$ theory and construct gravity towers for $N=2$ up to level $h=2$, and compare the resulting spectrum with the CFT\@. 
Section~\ref{sec:discussion} is devoted to discussion.
In Appendix~\ref{app:expl_states}, we write down the explicit forms of the states used in the main text.
In Appendix~\ref{app:chars}, we present various characters, including a generalization of the contracted large $\cN=4$ characters that include fugacities for outer automorphism charges.

\section{CFT background}
\label{sec:background}

We study the space of $1/4$-BPS states of the D1-D5 CFT, working from the symmetric orbifold $\mathrm{Sym}^N(T^4)$ description, which can represent either excitations in supergravity, stringy states, or black hole microstates. While we do not attempt to introduce all relevant background concepts here, introductions to the D1-D5 system and its holographically dual CFT can be found, for example, in~\cite{Avery:2010qw,David:2002wn,Hughes:2025tdy}. In particular, our conventions in this paper are exactly the same as those in \cite{Hughes:2026qqn}.

The seed theory used in constructing $\mathrm{Sym}^N(T^4)$ is the supersymmetric sigma model on~$T^4$, containing 4 free bosons $X^{\Ad A}$, 4 left-moving free fermions $\psi^{\alpha\Ad}$, and 4 right-moving free fermions $\tilde{\psi}^{\alphad \Ad}$. This seed theory, defined on a spatial $S^1$, has, central charge $c_0=6$, $\cN=(4,4)$ supersymmetry, as well as $SU(2)_L\times SU(2)_R$ R-symmetry, for which $\alpha=\pm$, $\alphad={\dot{\pm}}$ are doublet indices. Likewise, the $A=1,2$ and $\Ad= \dot{1},\dot{2}$ indices are doublet indices for $SU(2)_1\times SU(2)_2$ which, while not a symmetry of the theory,\footnote{The $SU(2)_1\times SU(2)_2$ is a broken symmetry due to states carrying non-trivial momentum and winding on the torus. In this paper we will only discuss the CFT's BPS sector, contained within the zero momentum and winding sector, and so these $SU(2)$ will be used extensively to organize states.} serve to organize the free bosons which parameterize the $T^4$ target space. In this paper we choose to work in the Neveu-Schwarz (NS) sector of the CFT, in which the fermion fields are anti-periodic on the~$S^1$.

In the symmetric orbifold, the Hilbert space decomposes into twist sectors $H_{\{n_k\}}$ labeled by integer partitions of $N$ (\textit{i.e.}, $\sum_{k=1}^{N}k n_k=N$). In turn, each twist sector can be described by a collection of $n$ ``strands'' (where $\sum_{k=1}^{N}n_k=n$): on a strand of length $k$, the Hilbert space $H_{(k)}$ is that of the seed theory defined on a $k$-wound spatial $S^1$ with the restriction $L_0-\tilde{L}_0\in\mathbb{Z}$. If, for a given twist sector, there are multiple strands of the same length (\textit{i.e.}, $n_k>1$) only the combination invariant under permutations of strands is kept. This decomposition of the full Hilbert space of $\mathrm{Sym}^N(T^4)$ can be summarized as
\begin{equation} \label{eq:Hilbert_space_N}
    H\big(\mathrm{Sym}^N(T^4)\big) = \bigoplus_{\{n_k\}} \bigotimes_{k=1}^{N} \Bigl(\underbrace{H_{(k)}\otimes\cdots\otimes H_{(k)}}_{n_k}\Bigl)_{\text{$S_{n_k}$-inv}} \ ,
\end{equation}
where $(\,\cdot\,)_{\text{$S_{n}$-inv}}$ denotes the $S_n$-invariant subspace of an $n$-fold tensor product space.

On the $I$th strand, whose length is $k_I$, the expansion modes of the fields $\pd X^{\Ad A}$, $\psi^{\alpha \Ad}$, $\tilde{\pd} X^{\Ad A}$, $\tilde{\psi}^{\ald \Ad}$ are
\begin{align} \label{eq.alpha_psi_modes_k}
    \alpha^{\Ad A[I]}_{s} \ ,\ \ \psi^{\alpha \Ad[I]}_{s} \ ,\ \ \tilde{\alpha}^{\Ad A[I]}_{s}\ ,\ \ \tilde{\psi}^{\ald \Ad[I]}_s \ ,
\end{align}
where $[I]$ denotes the strand and $s=\frac{r}{k_I}$ with $r\in\mathbb{Z}$ for bosons and $r\in\mathbb{Z}+\frac12$ for fermions. The generators of the superconformal algebra of $\mathrm{Sym}^N(T^4)$ are the diagonal---or ``total''\footnote{%
Generally, total generators, denoted by $(\mathrm{T})$, are given in terms of the individual strand modes by
\begin{equation}
    \mathcal{O}^{(\mathrm{T})}_r = \sum_{I=1}^{n} \mathcal{O}^{[I]}_r \ .
\end{equation}}%
---generators of the $N$ copies of the seed theory symmetry algebra, namely
\begin{equation} \label{eq.currentModes}
    \Big\{ L_{r}^{(\mathrm{T})}\ ,\ J^{\pm,3(\mathrm{T})}_{r}\ ,\ G^{\alpha A(\mathrm{T})}_{r+\half}\ ,\ \alpha^{\Ad A(\mathrm{T})}_{r} \ ,\ \ \psi^{\alpha \Ad(\mathrm{T})}_{r+\half}\Big\} \ ,
    %\ \ ,\ \ \Big\{ \tilde{L}_{r}^{(\mathrm{T})}\ ,\ \tilde{J}^{\dot{\pm},3(\mathrm{T})}_{r}\ ,\ \tilde{G}^{\ald A(\mathrm{T})}_{r}\Big\} \ ,
\end{equation}
for $r\in\mathbb{Z}$ and which satisfy a 2d contracted large $\cN=4$ superconformal algebra (found, for example, in \cite[App. A]{Hughes:2026qqn}).   
The analogous generators for the right-moving sector are denoted with a tilde. 
We will refer to the generators \eqref{eq.currentModes} collectively as ``affine generators'' in the following sections.  These affine generators remain symmetry generators even away from the orbifold point, unlike the modes on individual strands.

Affine primary states are those annihilated by all positive affine modes. Chiral (left-moving) states are those saturating the unitarity bound $h=m=j$ (where $h$ and $m$ are eigenvalues of $L_0$ and $J^3_0$ respectively, and we use $j$ for the eigenvalue of the $SU(2)_L$ Casimir), while the analogous right-moving states satisfy $\tilde{h}=\mt$. On a single strand of length $k=1,\dots,N$, they are given by
\begin{subequations} \label{eq.chiral_primaries}
\begin{align}
    \text{Left-moving}:\qquad |\pm\rangle_k\ &: \ h=m=\frac{k\pm1}{2}\quad ,\quad |\Ad\rangle_k \ : \ h=m=\frac{k}{2} \ , \label{eq.chiral_primaries_L}\\
    \text{Right-moving}:\qquad |\dot{\pm}\rangle_k\ &: \ \tilde{h}=\mt=\frac{k\pm1}{2}\quad ,\quad |\Ad\rangle_k \ : \ \tilde{h}=\mt=\frac{k}{2} \ . \label{eq.chiral_primaries_R}
\end{align}
\end{subequations}
The unique NS vacuum, with $h=m=\tilde{h}=\mt=0$, is in the untwisted sector (the twist sector satisfying $n_k=N\delta_{k,1}$) and is given by $\bigotimes_{I=1}^{N}\ket{-}_1^{[I]}\ket*{\dot{-}}_1^{[I]}$.

This paper focuses on the $1/4$-BPS sector of the CFT\@. In the symmetric orbifold description, these are simply states arbitrarily excited on each strand by left-moving modes \eqref{eq.alpha_psi_modes_k} and whose right-moving part is chiral. These BPS states satisfy the shortening condition $\tilde{G}^{\dot{+}A}_{-\frac12}\ket*{\rm{BPS}}=0$. 

Supergraviton states, or graviton states for short, in the symmetric orbifold are a special class of $1/4$-BPS states, for which the left-moving excitations on each strand are from ``individual global modes'' on that strand---the modes generating the $su(1,1|2)$ anomaly-free subalgebra: \textit{i.e.}, the set
$\{ L_{\pm1,0}\,,\, G^{\alpha A}_{\pm\frac12}\,,\, J^{\pm,3}_0 \}$.
Explicitly, the most general multi-strand supergraviton state is a product of components of the form
\begin{equation} \label{eq:graviton_states_def}
\begin{aligned}
    &\prod_{\psi,r,m} \Big[\big(L_{-1}\big)^r\big(J^-_{0}\big)^m|\psi\rangle\Big]^{n_{\psi,r,m}} \ , \\
    &\prod_{\psi,r,m} \Big[\big(L_{-1}\big)^r\big(J^-_{0}\big)^m G^{-A}_{-\frac12}|\psi\rangle\Big]^{n^{'}_{\psi,r,m}} \ ,\\
    &\prod_{\psi,r,m} \Big[\big(L_{-1}\big)^r\big(J^-_{0}\big)^m \big( G^{-1}_{-\frac12}G^{-2}_{-\frac12} - \tfrac1{2h}L_{-1}J^{-}_0\big)|\psi\rangle\Big]^{n^{''}_{\psi,r,m}} \ ,
\end{aligned}
\end{equation}
where here $\ket{\psi}$ refers to the set of (left times right) single-strand chiral states \eqref{eq.chiral_primaries} of left-conformal weight $h$, and $n_{\psi,r,m},n^{'}_{\psi,r,m},n^{''}_{\psi,r,m}\in\mathbb{Z}_+$. Since we focus on the $N=2$ theory in this paper, states can only have 1 or 2 strands and so $n_{\psi,r,m},n^{'}_{\psi,r,m},n^{''}_{\psi,r,m}\leq2$.
Graviton states can be organized into total global multiples, \textit{i.e.}\ representations of the $su(1,1|2)$ algebra generated by the total global modes 
\begin{equation} \label{eq:total_lobal_modes}
    \Big\{ L_{\pm1,0}^{\rm (T)}\,,\, G^{\alpha A{\rm (T)}}_{\pm\frac12}\,,\, J^{\pm,3{\rm (T)}}_0 \Big\} \ .
\end{equation}
Henceforth, we will refer to total global multiplets simply as global multiplets.  Global multiplets can be short or long, and their characters are found in \eqref{eq.SU112_short_def} and  \eqref{eq.SU112_long_def}.

Singletons are boundary diffeomorphisms in the bulk~\cite{Brown:1986nw} and correspond in the CFT to total affine generators \eqref{eq.currentModes}.  Acting with affine generators on graviton states generally produces states that are no longer gravitons.\footnote{There are, however, exceptions, which complicate the construction of $H_0^{\rm gravity}$ from  $H_0^{\rm graviton}$.  For example, $J^+_{-(n+1)}\ket{-}_1={1\over n!}L_{-1}^n\ket{+}_1$, $n\in\mathbb{Z}_{\ge 0}$.}  We can act with affine generators multiple times on a graviton state. In this way, the action of affine generators extends the graviton Hilbert space $H_0^{\rm graviton}$ into the gravity Hilbert space $H_0^{\rm gravity}$, which is closed under their action. The states in $H_0^{\rm gravity}$ decompose into affine multiplets, \textit{i.e.}\ representations of the $\cN=4$ superconformal algebra.  Affine multiplets can be short or long, and their characters are found in \eqref{Phi(s)_N=4} and \eqref{Phi(l)_N=4}.

Due to the phenomena of multiplet recombination, deforming $\mathrm{Sym}^N(T^4)$ by a twisted-sector exactly marginal operator leads to many free-theory short multiplets of $1/4$-BPS states combining into long multiplets of the interacting theory: these states gain anomalous dimensions and are ``lifted'' above the BPS bound. This lifting of states has been studied, for example in \cite{Gava:2002xb,Hampton:2018ygz,Guo:2022ifr,Guo:2020gxm,Guo:2019pzk,Gaberdiel:2023lco,Fabri:2025rok,Benjamin:2021zkn}, and has been recently made sharper by the ideas of fortuity~\cite{Chang:2025rqy,Chang:2025wgo}. Under the program of fortuity, unlifted free-theory BPS states fall into two types: monotone states, which are unlifted for all values of $N$ and so have a smooth large $N$ limit, correspond to states visible in supergravity; fortuitous states, which are unlifted only for a finite range of $N$ and so have no smooth large $N$ limit, correspond to typical BPS black hole microstates~\cite{Chang:2024zqi}. In Section~\ref{sec:method} we discuss a method for explicitly constructing the affine multiplets in the monotone sector.

In \cite{Hughes:2025oxu,Hughes:2026qqn} a description of the $1/4$-BPS Hilbert space of $\mathrm{Sym}^N(T^4)$ based on the Schur-Weyl duality \cite{Fulton:1991} was put forward. The general idea behind this description is that, while the physical BPS states should be invariant under permutations of strands, the left- and right-moving parts of states can individually transform non-trivially.\footnote{This is quite analogous to standard constructions of the wavefunction of a system of spin-1/2 particles in quantum mechanics; the orbital and spin wavefunctions can have non-trivial transformation under the exchange of particles, as long as the total wavefunction is antisymmetric.} In a twist sector involving $n$ strands, the left- and right-moving BPS Hilbert spaces decompose into ``symmetry sectors'', $V_{\lam}$ and $\tilde{V}_{\lam}$ respectively, labeled by Young diagrams $\lam$ with $n$ boxes. For overall $S_n$ invariance, the left- and right-moving parts of states have to be in the same symmetry sector.

In this paper we will only deal with the $N=2$ theory, for which the BPS Hilbert space admits the Schur-Weyl decomposition \cite{Hughes:2026qqn}
\begin{equation} \label{eq:H_N=2_SW}
    H_2^{\mathrm{CFT}} \equiv H\big(\mathrm{Sym}^2(T^4)\big)\big|_{\mathrm{BPS}} = \big(V_{\YTscriptsize\ydiagram{2}}\otimes\Vt_{\ydiagram{2}}\big) \oplus \big(V_{\ydiagram{1,1}}\otimes\Vt_{\ydiagram{1,1}}\big) \oplus \big(V_{\ydiagram{1}}\otimes\Vt_{\ydiagram{1}}\big) \ ,
\end{equation}
where the $\lam=\CYon\YTnormalsize\ydiagram{2}$ and $\ydiagram{1,1}$ sectors contain states with two strands of length 1 that transform symmetrically and anti-symmetrically, respectively, under the exchange of strands. The twisted-sector states are contained within the $\lam=\ydiagram{1}$ sector. Of central importance in this paper, the supergraviton subsector of the free CFT spectrum also admits the symmetry-sector decomposition \eqref{eq:H_N=2_SW}. The right-moving BPS Hilbert spaces $\tilde{V}_{\lam}$ are common to both the full BPS CFT spectrum and the supergraviton sector, coming just from right-chiral states. These are given as follows:
\begin{subequations} \label{eq:right-moving_states}
\begin{align}
    \tilde{V}_{\YTscriptsize\ydiagram{2}} &= \mathrm{span}\Big\{ \ket*{\dot{-}}_1^{[1]}\ket*{\dot{-}}_1^{[2]}\,,\ \ket*{\dot{-}}_1^{[1]}\ket*{\dot{A}}_1^{[2]} + \ket*{\dot{A}}_1^{[1]}\ket*{\dot{-}}_1^{[2]}\,,\ \ket*{\dot{-}}_1^{[1]}\ket*{\dot{+}}_1^{[2]} + \ket*{\dot{+}}_1^{[1]}\ket*{\dot{-}}_1^{[2]}\,, \nonumber\\
    &\qquad\qquad \ep_{\Ad\Bd}\big(\ket*{\dot{A}}_1^{[1]}\ket*{\dot{B}}_1^{[2]} - \ket*{\dot{B}}_1^{[1]}\ket*{\dot{A}}_1^{[2]}\big)\,,\ \ket*{\dot{+}}_1^{[1]}\ket*{\dot{A}}_1^{[2]} + \ket*{\dot{A}}_1^{[1]}\ket*{\dot{+}}_1^{[2]}\,,\ \ket*{\dot{+}}_1^{[1]}\ket*{\dot{+}}_1^{[2]} \Big\} \ , \label{eq:right-moving_states_2}\\
    \tilde{V}_{\ydiagram{1,1}} &= \mathrm{span}\Big\{ \ket*{\dot{-}}_1^{[1]}\ket*{\dot{A}}_1^{[2]} - \ket*{\dot{A}}_1^{[1]}\ket*{\dot{-}}_1^{[2]}\,,\ \ket*{\dot{-}}_1^{[1]}\ket*{\dot{+}}_1^{[2]} - \ket*{\dot{+}}_1^{[1]}\ket*{\dot{-}}_1^{[2]}\,, \nonumber\\
    &\qquad\qquad \sigma^{a}_{\Ad\Bd}\big(\ket*{\dot{A}}_1^{[1]}\ket*{\dot{B}}_1^{[2]} + \ket*{\dot{B}}_1^{[1]}\ket*{\dot{A}}_1^{[2]}\big)\,,\ \ket*{\dot{+}}_1^{[1]}\ket*{\dot{A}}_1^{[2]} - \ket*{\dot{A}}_1^{[1]}\ket*{\dot{+}}_1^{[2]} \Big\} \ , \label{eq:right-moving_states_1,1}\\
    \tilde{V}_{\ydiagram{1}} &= \mathrm{span}\Big\{ \ket*{\dot{-}}_2^{[1]} \,,\ \ket*{\dot{A}}_2^{[1]}\,,\ \ket*{\dot{+}}_2^{[1]} \Big\} \ , \label{eq:right-moving_states_1}
\end{align}
\end{subequations}
where our conventions for the $SU(2)$ singlet and triplet projectors $\epsilon_{\Ad\Bd},\sigma^a_{\Ad\Bd}$ can be found in \cite[Eq. A.5]{Hughes:2026qqn}.

From the ``Schur-Weyl form'' of the BPS Hilbert space in \eqref{eq:H_N=2_SW}, the BPS partition function for $\mathrm{Sym}^2(T^4)$ takes the form~\cite{Hughes:2026qqn}
\begin{align}
    Z_{N=2}(q,y,\eta,\dot{\eta},\yt) &= \Tr_{H^{\rm BPS}_{\rm NS}} \!\big[ (-1)^{F} q^{L_0}y^{2J^3_0} \eta^{2K^3_{1,0}} \dot{\eta}^{2K^3_{2,0}} \yt^{2\tilde{J}^3_0}\big] \nonumber\\
    &= S_{\YTscriptsize\ydiagram{2}}(q,y,\eta,\dot{\eta}) \tilde{S}_{\YTscriptsize\ydiagram{2}}(\yt) + S_{\YTscriptsize\ydiagram{1,1}}(q,y,\eta,\dot{\eta}) \tilde{S}_{\YTscriptsize\ydiagram{1,1}}(\yt) + S_{\YTscriptsize\ydiagram{1}}(q,y,\eta,\dot{\eta}) \tilde{S}_{\YTscriptsize\ydiagram{1}}(\yt) \ ,\label{eq:SW_PF}
\end{align}
where $S_{\lam}$ and $\tilde{S}_{\lam}$ are the contributions respectively from the left- and right-moving states in the symmetry sector labelled by $\lam$. We have chosen here to include the fugacities $\eta$,$\dot{\eta}$ for the charges of left-moving states under $SU(2)_1$ and $SU(2)_2$ respectively. In \eqref{eq:SW_PF}, $K^3_{1,0}$ and $K^3_{2,0}$ are zero modes of the third component generators of $SU(2)_1$ and $SU(2)_2$. These extra fugacities are useful for keeping track of individual states, as we will demonstrate in the current paper.
The form of the expression \eqref{eq:SW_PF} is equally valid for the full BPS spectrum of the symmetric orbifold theory and a subsector, such as the supergraviton sector; in each case the appropriate BPS Hilbert space $H^{\rm BPS}_{\rm NS}$ should be traced over.

Being the characters of the spaces $V_{\lam}$, the $S_{\lam}$ in \eqref{eq:SW_PF} are given by the Schur functions
\begin{subequations} \label{eq:Schur_poly_defs}
    \begin{align}
        S_{\YTscriptsize\ydiagram{2}}(q,y,\eta,\dot{\eta}) &= \frac12\Big[z_1(q,y,\eta,\dot{\eta})^2 + z_1(q^2,y^2,\eta^2,\dot{\eta}^2)\Big] \ , \label{eq:S_2_def}\\
        S_{\YTscriptsize\ydiagram{1,1}}(q,y,\eta,\dot{\eta}) &= \frac12\Big[z_1(q,y,\eta,\dot{\eta})^2 - z_1(q^2,y^2,\eta^2,\dot{\eta}^2)\Big] \ ,\label{eq:S_1,1_def}\\
        S_{\YTscriptsize\ydiagram{1}}(q,y,\eta,\dot{\eta}) &= z_2(q,y,\eta,\dot{\eta}) \ ,\label{eq:S_1_def}
    \end{align}
\end{subequations}
where $z_k$ is the left-moving contribution to the partition function from states on a strand of length $k$ discussed below. For the right-moving spaces \eqref{eq:right-moving_states}, the Schur functions $\tilde{S}_{\lam}$ are defined analogously to \eqref{eq:Schur_poly_defs}, in terms of the right-moving single-strand contributions $\tilde{z}_k(\yt)=\yt^{k-1}(1-\yt)^2$.

For the left-moving CFT spectrum, the single-strand partition functions $z_k^{\mathrm{CFT}}$ are given in terms of the expansion coefficients $c(x;\eta,\etad)$ of the seed theory partition function (defined in \eqref{eq:seed_z(q,y)}), by
\begin{equation} \label{eq:zk_CFT}
    z_k^{\mathrm{CFT}}(q,y,\eta,\dot{\eta}) = \sum_{r,\ell} c(4kr-\ell^2;\eta,\etad) q^r y^{\ell} \ .
\end{equation}
For the supergraviton sector, the left-moving single-strand partition function $z_k^{\mathrm{graviton}}$ is given by
\begin{equation}
    z_k^{\mathrm{graviton}}(q,y,\eta,\etad) = \phi_{\frac{k-1}{2}}(q,y,\eta) - \dot{\chi}_{\frac12} \phi_{\frac{k}{2}}(q,y,\eta) + \phi_{\frac{k+1}{2}}(q,y,\eta) \ ,
\end{equation}
where $\phi_{j}$ are short characters of $SU(1,1|2)$, generalized to include the $SU(2)_1$ fugacity $\eta$, while $\dot{\chi}_j$ are characters of $SU(2)_2$, as defined in Appendix~\ref{app:chars}.

As described in section~\ref{sec:intro}, the goal of this paper is to understand the multiplet structure of the BPS Hilbert space \eqref{eq:H_N=2_SW}. As such, we will construct the explicit affine primary states up to level $h=2$ that are captured by the character decomposition of the partition functions \eqref{eq:SW_PF}. The character decompositions of the various symmetry sectors \eqref{eq:Schur_poly_defs} of the free CFT partition function are given by
\begin{subequations} \label{eq:S_CFT_exp_refined}
    \begin{align}
        S^{\mathrm{CFT}}_{\YTscriptsize\ydiagram{2}} &= \mathbf{\Phi}_{0} + \dot{\chi}_1 \mathbf{\Phi}_{0,1} + \mathbf{\Phi}_{0,2} + \big(\dot{\chi}_1 + \chi_{1} \dot{\chi}_{2} \big)\mathbf{\Phi}_{0,3} + \cdots \ ,\label{eq:S_CFT_2_exp_refined}\\
        S^{\mathrm{CFT}}_{\YTscriptsize\ydiagram{1}} &= \mathbf{\Phi}_{\frac12} - \chi_{\frac12} \dot{\chi}_{1} \mathbf{\Phi}_{0,1} - \big(\chi_{\frac12} + \chi_{\frac12} \dot{\chi}_{1} + \chi_{\frac32} \dot{\chi}_{2} \big) \mathbf{\Phi}_{0,2} \nonumber\\
        &\quad - \Big(\chi_{\frac12} + \big(2\chi_{\frac12}+\chi_{\frac32}\big)\dot{\chi}_{1} + \big(\chi_{\frac12}+\chi_{\frac32}\big)\dot{\chi}_{2} + \chi_{\frac52} \dot{\chi}_{3} \Big) \mathbf{\Phi}_{0,3} + \cdots \ ,\label{eq:S_CFT_1_exp_refined}\\
        S^{\mathrm{CFT}}_{\YTscriptsize\ydiagram{1,1}} &= -\dot{\chi}_{\frac12} \mathbf{\Phi}_{\frac12} + \chi_{\frac12} \dot{\chi}_{\frac32} \mathbf{\Phi}_{0,2} + \chi_{\frac12} \dot{\chi}_{\frac12} \mathbf{\Phi}_{0,3} + \cdots \ , \label{eq:S_CFT_1,1_exp_refined}
    \end{align}
\end{subequations}
where $\mathbf{\Phi}_{j}(q,y,\eta,\etad)$ and $\mathbf{\Phi}_{j,h}(q,y,\eta,\etad)$ are the contracted large $\cN=4$ characters, generalized to include $SU(2)_1$ and $SU(2)_2$ fugacities.
Their expressions for general $N$ are given in Appendix~\ref{app:chars}; in the present paper, we are specializing to $N=2$. The $\chi_j$ are the characters of $SU(2)_1$. The characters $\chi_j,\dot\chi_j$ help us find explicit affine primary states.  In the following sections we seek to understand the nature of states contributing to the above character decompositions.

\section{How to find gravity affine multiplets} \label{sec:method}

Here we present a general procedure to find the full set of supergraviton states dressed with singleton excitations.  
For this purpose, we can restrict to a fixed symmetry sector labeled by $\lambda$, because adding singletons---being total modes which are invariant under $S_n$---does not change the symmetry sector.  This means that we can forget about the right-moving sector and focus on the left-moving sector with the specified symmetry $\lambda$ under strand exchange, which we shall do below.
    
\bigskip

Let $H^{\rm graviton}_0$ denote the space of all BPS graviton states and let $\cA$ denote the algebra of total affine generators representing all possible singleton excitations.  The subscript $0$ on $H^{\rm graviton}_0$ means that this space is defined in the free symmetric orbifold CFT\@.  As explained above, we are focusing on the left-moving sector with symmetry type $\lambda$ and thus, strictly speaking, we must write it as $H^{\rm graviton}_{0,N,\lambda}$, but we are suppressing these labels for simplicity.  Then the space of BPS states in the gravity sector is
\begin{align}
 H^{\rm gravity}_0=\cA\cdot H^{\rm graviton}_0.
 \label{H_gty=A.H_gton}
\end{align}
The product in \eqref{H_gty=A.H_gton} contains redundancy, in that the same state may be obtained from different affine generators acting on different graviton states.
We can organize states in $H^{\rm gravity}_0$ into multiplets of the total affine algebra.\footnote{Such a decomposition is not possible for $H^{\rm graviton}_0$, which only admits a decomposition into multiplets of the anomaly-free subalgebra, \textit{i.e.}\ into global multiplets.}  We will refer to those affine multiplets as \textit{gravity towers}. Our goal is to start from from $H^{\rm graviton}_0$ and find all gravity towers  without overcounting, taking into account the redundancy.

Let us denote the gravity towers in \eqref{H_gty=A.H_gton} by ${\bf a}^{(t)}$, where $t=1,2,\dots,n$.   The states in ${\bf a}^{(t)}$ are generated by the action of affine raising generators on an affine primary state $\ket*{a^{(t)}_{J_t,H_t}}$ of charges $(J_t,H_t)$, which is annihilated by all affine lowering generators.  We refer to $(J_t,H_t)$ as the affine-primary charges of the tower ${\bf a}^{(t)}$.  We sometimes denote the affine primary by ${\bf a}^{(t)}_{J_t,H_t}$, displaying its affine-primary charges.
Different gravity towers are mutually orthogonal: if $t\neq t'$, then every state in ${\bf a}^{(t)}$ is orthogonal to every state in ${\bf a}^{(t')}$.
As affine multiplets, the gravity tower ${\bf a}^{(t)}_{J_t,H_t}$ is either short ($J_t=H_t$) or long ($J_t<H_t$).    The possible ranges for $J_t$ are 
\begin{align}
J_t=\begin{cases}
\ 0,\tfrac12,\dots,\tfrac{N}{2}-\tfrac12 & \text{for short towers,}\\[1.2ex]
\ 0,\tfrac12,\dots,\tfrac{N}{2}-1 &\text{for long towers.}
\end{cases}
\label{J_t_range}
\end{align}

\begin{figure}[tb]
\begin{center}
\begin{tikzpicture}[scale=0.9]
\footnotesize
\begin{scope}
\clip (-5,-1) rectangle (5,6.5);
\draw[fill=black!5!white,dashed] (-4.5,7.5) -- (-3,3) -- (0,0) -- (3,3) -- (4.5,7.5) -- cycle;
\draw[color=blue,fill=blue!20!white] (-4.5,8.5) -- (-3.5,5.5) -- (-2.5,3.5) -- (-0.5,1.5) -- (0.5,1.5) -- (2.5,3.5) -- (3.5,5.5) -- (4.5,8.5) -- cycle;
%\draw[fill=blue] (-3.5,5.5) circle (0.05);
%\draw[fill=blue] (-2.5,3.5) circle (0.05);
%\draw[fill=blue] (-0.5,1.5) circle (0.05);
%\draw[fill=blue] (+3.5,5.5) circle (0.05);
%\draw[fill=blue] (+2.5,3.5) circle (0.05);
\draw[fill=blue] (+0.5,1.5) circle (0.075);
\end{scope}
\draw[dotted] (3,3) -- (3,0) node [below] {$N$};
\draw[dotted] (3,3) -- (0,3) node [left,xshift=2] {$N$};
\node[color=blue] at (0,1.5) [above left,xshift=3,yshift=-2] {$H$};
\draw[dotted,color=blue] (-0.5,1.5) -- (-0.5,0) node [below,xshift=-7] {$-J$};
\draw[dotted,color=blue] (0.5,1.5) -- (0.5,0) node [below] {$J$};
\draw[densely dotted,color=blue] (2.5,3.5) -- (2.5,0) node [above left,xshift=2,yshift=-2] {$N\!-\!J$};
%\draw[densely dotted,color=blue] (2.5,3.5) -- (2.5,0);
%\draw[latex-,color=blue] (2.5,0) -- +(0,-1) node [below,xshift=-5,yshift=2] {$N\!-\!J$};
\draw[densely dotted,color=blue] (3.5,5.5) -- (3.5,0) node [above right,xshift=-2,yshift=-2] {$N\!+\!J$};
%\draw[densely dotted,color=blue] (3.5,5.5) -- (3.5,0);
%\draw[latex-,color=blue] (3.5,0) -- +(0,-1) node [below,xshift=5,yshift=2] {$N\!+\!J$};
\draw[densely dotted,color=blue] (2.5,3.5) -- (0,3.5) node [left,xshift=2] {$N\!+\!H\!-\!2J$};
\draw[densely dotted,color=blue] (3.5,5.5) -- (0,5.5) node [left,xshift=2] {$N\!+\!H\!+\!2J$};
% axes
\draw[-latex] (-5,0) -- (5,0) node [right] {$j$};
\draw[-latex] (0,-1) -- (0,7) node [above] {$h$};
\end{tikzpicture}
  \caption{\sl The structure of the affine tower ${\bf a}_{J,H}$ on the $j$-$h$ plane. The tower ${\bf a}_{J,H}$ contains global multiplets ${\bf s}_{j,h}$ with $(j,h)$ in the blue shaded region.  The dot indicates the ``base'' of the tower where the affine primary is.  States of the CFT exist only in the gray shaded region bounded from below by the unitarity bound shown as the black dashed lines.
  %For the same value of $H$, an affine tower with a larger value of $J$ covers a wider region and is regarded as smaller.
  \label{fig:structure_affine_tower}}
\end{center}
\end{figure}
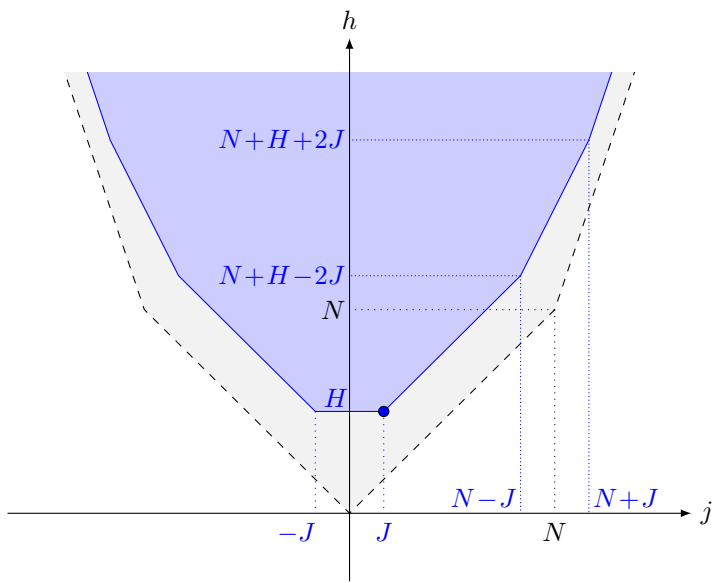

We can decompose a gravity tower ${\bf a}^{(t)}$ into global multiplets as
\begin{align}
{\bf a}^{(t)}_{J_t,H_t}=\bigoplus_{j,h,i} {\bf s}^{(t)i}_{j,h},
\label{a^t=sum_g^t}
\end{align}
where ${\bf s}^{(t)i}_{j,h}$ denotes a global multiplet built on a global primary of charges ($j,h$), which may again be short ($j=h$) or long ($j<h$).  We refer to ${\bf s}^{(t)i}_{j,h}$ as \textit{singleton global multiplets}.  The index $i$ distinguishes global multiplets with the same $(j,h)$ when there is more than one such multiplet present.
We refer to $(j,h)$ as the global-primary charges of the singleton global multiplet ${\bf s}^{(t)i}_{j,h}$.  Given $(J_t,H_t)$, the range of $(j,h)$ in the sum is restricted; see Figure \ref{fig:structure_affine_tower}. 
%For example, if $N=2$, the affine multiplet ${\bf a}_{0,0}$ has the 

Let us introduce an ordering on charges, applicable both to affine-primary charges and global-primary charges, as follows: we define 
\begin{align}
    (J,H)<(J',H')  \qquad \text{if $H<H'$,~~ or ~~ if $H=H'$ and $J>J'$}.
    \label{eq:(J,H)_ordering}
\end{align}
Namely, the ordering is by increasing $H$, and for fixed $H$, by decreasing $J$.
This ordering has the following key property: a global multiplet with global-primary charges $(j,h)$ does not occur in an affine tower ${\bf a}_{J,H}$ if $(j,h)<(J,H)$.

Let us assume that the gravity towers ${\bf a}^{(t)}_{J_t,H_t}$ are ordered so that
\begin{align}
    (J_1,H_1)< (J_2,H_2) < \cdots~.
    \label{(j,h)_ordering}
\end{align}
More generally, there can be more than one gravity tower with the same affine-primary charges; for example, gravity towers may come in non-trivial $su(2)_{1}$ and $su(2)_{2}$ multiplets.  We will discuss this more general case later.

Each gravity tower must be related to pure gravitons without singleton dressing, in order for it to be generated by the action of total affine modes on pure gravitons as in \eqref{H_gty=A.H_gton}.  Roughly speaking, each gravity tower ${\bf a}^{(t)}$ must contain at least one global multiplet ${\bf s}_{j,h}^{(t)i}$ of pure gravitons.  However, the general situation is more complicated: the pure graviton states need not coincide with a single global multiplet from one tower, but can instead be a linear combination of global multiplets from different towers with the same $(j,h)$. Therefore, the general relation takes the form of a system of equations labeled by $p=1,\dots,m$ where $m\ge n$:
\begin{align}
 \sum_{t,i} c^{(t)i}_p\, {\bf s}^{(t)i}_{j_p,h_p} = {\bf g}_{j_p,h_p}
.
\label{gjkh9Aug25}
\end{align}
Here the global multiplet ${\bf s}^{(t)i}_{j_p,h_p}$ belongs to the $t$-th gravity tower ${\bf a}^{(t)}$ (in the sense of \eqref{a^t=sum_g^t}), while ${\bf g}_{j_p,h_p}$ is a global multiplet made of pure gravitons, which we call \textit{graviton global multiplets}.\footnote{The notions of singleton global multiplet and gravity global multiplet are not mutually exclusive; a singleton global multiplet may simultaneously be a gravity global multiplet.}  The coefficients $c^{(t)i}_p$ are not generally integers. The $p$-th relation relates singleton global multiplets with the same charges $(j_p,h_p)$ from different gravity towers ${\bf a}^{(t)}$.  Note that the sum is finite because only towers $t$ satisfying $(J_t,H_t)\le (j_p,h_p)$ can contribute.

Given a relation of the form \eqref{gjkh9Aug25} among singleton global multiplets with $(j_p,h_p)$, we can derive relations among singleton global multiplets with smaller\footnote{By ``smaller'', we mean in the sense of the ordering \eqref{(j,h)_ordering}.} values of $(j,h)$, by acting with affine lowering generators.  Even after doing this, the right-hand side remains gravitons for the following reason.  Graviton states are made of global generators $L_{-1}$, $G_{-\half}^{\alpha \Ad}$, and $J_0^-$ acting on chiral states.  Suppose that a given graviton state involves $a$ factors of $L_{-1}$ and $b$ factors of $G_{-\half}$.  When we act on this state with an affine generator, \textit{e.g.}\ $L_{r>0}$, we must have $r\le a+b/2$ so that $L_r$ does not annihilate the state.  As $L_r$ is commuted through $L_{-1}$ and $G_{-1/2}$, its mode number decreases and eventually enters the range of the global (anomaly-free) subalgebra.  After that, further commutation through $L_{-1}$ and $G_{-1/2}$ keeps it within the global subalgebra.  Therefore, the resulting state is again a graviton state.

This leads to a crucial observation about the relation \eqref{gjkh9Aug25}.  For given $p$, let $T_p$ denote the (finite) set of tower labels $t$ that appear in the sum \eqref{gjkh9Aug25}.  If $t\in T_p$, then $(J_t,H_t)\le (j_p,h_p)$, although not every $t$ satisfying $(J_t,H_t)\le (j_p,h_p)$ need belong to $T_p$.  Now, choose any tower $t'\in  T_p$ and act on \eqref{gjkh9Aug25}  with affine lowering generators (where we include $J^-_0$ among them), so as to lower the global multiplet ${\bf s}^{(t')i}_{j_p,h_p}$ in the sum to the lowest global multiplet ${\bf s}^{(t')}_{J_{t'},H_{t'}}$, \textit{i.e.}~the one containing the affine primary.  Because the lowest global multiplet in an affine tower is unique, ${\bf s}^{(t')}_{J_{t'},H_{t'}}$ has no index $i$.  This process of going down to the lowest global multiplet is always possible. Even after doing so, the left-hand side remains non-vanishing, because it is a sum of contributions from mutually orthogonal towers, and the contribution from $t'$ is non-vanishing.  It follows that the right-hand side, which is a graviton state, must also be non-vanishing.  Thus, the relation \eqref{gjkh9Aug25} implies another relation involving the lowest global multiplet of the tower $t'$.  Because $t'$ was an arbitrary element of $T_p$, we conclude that for every tower $t$ appearing in the sum in \eqref{gjkh9Aug25}, there exists a relation involving the lowest multiplet of that tower.  In other words, the existence of every gravity tower can be detected from its affine primary.

The above motivates the following process to solve the equations \eqref{gjkh9Aug25}.  
First, act on the equations with affine lowering generators to go down to the smallest possible affine-primary charges, $(J_1,H_1)$.  Because no towers other than $t=1$ contain global multiplets with these charges, we end up with a relation of the form
\begin{align}
 c^{(1)} {\bf s}^{(1)}_{J_1,H_1} &= {\bf g}_{J_1,H_1},
 \label{s^(1)=g_J1,H1}
\end{align}
where ${\bf s}^{(1)}_{J_1,H_1}$ is the lowest global multiplet of tower 1 and ${\bf g}_{J_1,H_1}$ is a pure-graviton global multiplet obtained by lowering ${\bf g}_{j_p,h_p}$. For consistency, ${\bf g}_{J_1,H_1}$ must be independent of $p$.
Because the global primary of ${\bf s}^{(1)}_{J_1,H_1}$ is nothing but the affine primary of tower 1, we conclude that tower 1's affine primary is a graviton state.
Next, we lower the relation \eqref{gjkh9Aug25} to $(J_2,H_2)$.  Because only towers $t=1,2$ contain global multiplets with these charges, we find a relation of the form
\begin{align}
 \sum_i c^{(1)i} {\bf s}^{(1)i}_{J_2,H_2}
 +c^{(2)} {\bf s}^{(2)}_{J_2,H_2}
 &= {\bf g}_{J_2,H_2},
\end{align}
where ${\bf g}_{J_2,H_2}$ is a pure-graviton global multiplet.  This can be written as
\begin{align}
 c^{(2)} {\bf s}^{(2)}_{J_2,H_2}
 &= {\bf g}_{J_2,H_2}- \sum_i c^{(1)i} {\bf s}^{(1)i}_{J_2,H_2}.
 \label{find_s^(2)}
\end{align}
Namely, we can express the affine primary of tower 2 in terms of a graviton state and a descendant in tower 1, which has already been found above.  
If we lower the relation \eqref{gjkh9Aug25} to $(J_3,H_3)$, we obtain a relation of the form
\begin{align}
 c^{(3)} {\bf g}^{(3)}_{J_3,H_3}
 &= {\bf g}_{J_3,H_3}
 - \sum_i c^{(1)i} {\bf s}^{(1)i}_{J_3,H_3}
 - \sum_i c^{(2)i} {\bf s}^{(2)i}_{J_3,H_3},
 \label{find_s^(3)}
\end{align}
which allows us to express the affine primary of tower 3 in terms of a graviton state and descendants in the towers already found.  By repeating this process, we can find all gravity towers.

In the above, we assumed that there is only one gravity tower with the same affine-primary charge $(J_t,H_t)$.  It is easy to relax this restriction.  Suppose that there are two gravity towers, $t=1$  and $t=1'$, with the same affine-primary charges $(J_1,H_1)$.  In that case, there must be not just one relation of the form \eqref{s^(1)=g_J1,H1}, but two independent relations which relate ${\bf s}^{(1)}_{J_1,H_1}$ and  ${\bf s}^{(1')}_{J_1,H_1}$ separately to two different pure-graviton global multiplets.  Otherwise, there would be only one point of connection to pure gravitons and  \eqref{H_gty=A.H_gton} would not generate two independent affine towers. Generalizing this, it is clear that all gravity towers can be identified from their affine primaries, even when multiple towers share the same affine-primary charges.

\bigskip
Based on the above discussion, the procedure for constructing all gravity towers from the graviton spectrum is as follows.  First we identify the graviton state (or possibly states) with the smallest charges $(J_1,H_1)$.  Such a state must be an affine primary and hence determines the first gravity tower, ${\bf a}^{(1)}_{J_1,H_1}$.  In practice, this step is straightforward. For example, for $N=2$ in the symmetric sector ($\YTnormalsize\lambda=\ydiagram{2}$\,), the graviton state with the smallest charges is $\ket*{-}_1^{[1]}\ket*{-}_1^{[2]}$, which is clearly an affine primary and therefore generates the first affine tower ${\bf a}^{(1)}_{0,0}$ built on it. 
Next, we move to the next value of $(J,H)$, increasing it by one step within the range \eqref{J_t_range}.  We then use \eqref{find_s^(2)} to determine if there exists an affine primary (or possibly primaries) with charges $(J,H)$. Concretely, we orthogonalize the graviton global-primary states with $(J,H)$ to the tower $t=1$.  If the resulting state is non-vanishing, then,  by \eqref{find_s^(2)}, it is the affine primary of the next tower ${\bf a}^{(2)}_{J_2,H_2}$, with $(J_2,H_2)=(J,H)$. If, on the other hand, the resulting state is zero, no affine primary exists at $(J,H)$ and we move on to the next value. 
We then further increase $(J,H)$ and orthogonalize the graviton global-primary states at that value to the $t=1$ and $t=2$ towers.   If the resulting state is non-vanishing, then,  by \eqref{find_s^(3)}, it is the affine primary of ${\bf a}^{(3)}_{J_3,H_3}$, with $(J_3,H_3)=(J,H)$.  Repeating this process, we can identify all gravity towers.

\subsubsection*{Differences from the method in \cite{Hughes:2025tdy}:}

In \cite{Hughes:2025tdy}, a method for finding gravity towers was proposed based on promoting gravity global multiplets to affine multiplets after taking into account the fact that some gravity global multiplets are related by affine (non-global) generators.  This method gives the correct generalized gravity spectrum up to certain levels as demonstrated in \cite{Hughes:2025tdy}.  However, above those levels, it fails to capture affine primaries that arise through mixing between gravity global primaries and affine descendants of affine primaries that have already been found, as in \eqref{find_s^(2)} and \eqref{find_s^(3)}.  The method proposed above overcomes this limitation and allows gravity towers to be identified at any level.
                
\section{Gravity affine towers up to level $h=2$} \label{sec:results}

\subsection{Explicit construction of gravity affine towers} \label{ssec:L_results}

% {\color{blue}
% \begin{itemize}
% \item
% This part  carries out the procedure laid out in the previous    
% section explicitly
% \item
%  Explain notation
% \item
%  Write down all chiral primaries up to level $h=2$. Write down the global primary expansion of the corresponding graviton partition function.  
%  \item Run the procedure from $h=0$ and up.
%  \item Don't write down explicit forms of the global primaries.  Just use their names like $\ket*{g^1_{0,1}}^a$.  Put the explicit forms in the appendix.
% \end{itemize}
%

In this section we carry out the procedure\footnote{These calculations were performed with the aid of Mathematica code.} laid out in the previous section and find gravity towers for the $N=2$ theory up to level $h=2$.  We work in the three symmetry sectors $\lambda=\ydiagram{2}$, $\ydiagram{1}$, $\ydiagram{1,1}$ separately.  For these symmetry sectors, the right-moving states are explicitly given in \eqref{eq:right-moving_states} and thus we will be focusing on the left-moving sector.

\YTnormalsize
\subsubsection{Untwisted symmetric sector ($\lambda=\ydiagram{2}\mkern1.5mu$)} \label{sssec:N=(1,1)_sym}

We start by discussing graviton global multiplets up to level $h=2$.  As reviewed in section~\ref{sec:background}, graviton states are constructed by tensoring single-graviton states, each of which is a global descendant of a single-strand chiral state \eqref{eq.chiral_primaries}. Thus, in the untwisted sector, graviton states are built on the following chiral states:
\begin{subequations}
\label{eq:gravitons_sym_sector}
    \begin{IEEEeqnarray}{cclll}
        \text{chiral states} &\qquad&  \text{graviton partition sum} &\qquad& \text{global character expansion}\notag\\
    \noalign{\vskip-2ex}
    \IEEEeqnarraymulticol{5}{c}{\makebox[16cm][l]{\rule{15cm}{0.4pt}}}\notag\\
    \noalign{\vskip-0.5ex}
        \ket*{-}_1 \ket*{-}_1 && 1 && \phi^{(0)}_{0}\label{eq:0}\\
        \ket*{-}_1 \ket*{\dot{A}}_1 && -\dot{\chi}_{\frac{1}{2}} \phi_{\frac{1}{2}} && -\dot{\chi}_{\frac{1}{2}} \phi^{(\frac{1}{2})}_{\frac{1}{2}}\label{eq:1/2}\\
        \ket*{\dot{A}}_1 \ket*{\dot{B}}_1 && \tfrac{1}{2} \bigl[ (-\dot{\chi}_\frac{1}{2} \phi_{\frac{1}{2}} )^2 + (-\dot{\chi}^{\ev{2}}_\frac{1}{2} \phi^{\ev{2}}_{\frac{1}{2}} ) \bigr] && \phi^{(1)}_{1} + (\dot{\chi}_1 \phi^{(1)}_{0,1} + \phi^{(1)}_{0,2} + \cdots )\label{eq:1}\\
        \ket*{-}_1 \ket*{+}_1 && \phi_{1} && \phi^{(1')}_{1}\label{eq:1'}\\
        \ket*{+}_1 \ket*{\dot{A}}_1 && \phi_1 (-\dot{\chi}_{\frac{1}{2}} \phi_{\frac{1}{2}}) && - \dot{\chi}_{\frac{1}{2}} \phi^{(\frac{3}{2})}_{\frac{3}{2}} - ( \dot{\chi}_{\frac{1}{2}} \phi^{(\frac{3}{2})}_{\frac{1}{2},\frac{3}{2}} + \dot{\chi}_{\frac{1}{2}} \phi^{(\frac{3}{2})}_{\frac{1}{2},\frac{5}{2}} + \cdots )         \label{eq:3/2} \\
        \ket*{+}_1 \ket*{+}_1 && \tfrac{1}{2} (\phi^{2}_1 + \phi^{\ev{2}}_1 ) && \phi^{(2)}_2 + (\phi^{(2)}_{0,2} + \phi^{(2)}_{1,3} + \cdots)\label{eq:2}
    \end{IEEEeqnarray}
\end{subequations}
Next to each chiral state, we have written the partition sum of the graviton states built on it, and its expansion into short and long global characters, $\phi_j$ and $\phi_{j,h}$. From this expansion, we can read off the graviton global multiplets and the corresponding global primaries present at $(h,j)$.  The $SU(2)_1,SU(2)_2$ characters $\chi_j,\dot{\chi}_j$ in front allow us to read off the $SU(2)_1\times SU(2)_2$ representations of these multiplets, which is quite useful in identifying the explicit forms of the global primaries.
In the expansion, we put the label $(h_0)$ on global characters as $\phi_j^{(h_0)},\phi_{j,h}^{(h_0)}$ to indicate the level $h_0$ of the chiral state from which the global multiplet originates, although, of course, the functional form of the character does not depend on the label.  
For level $h_0=1$, where there are two chiral states, we use two labels $(1)$ and $(1')$ to distinguish them.

For example, in \eqref{eq:2}, in the middle column, we have written the partition sum for the 2-graviton states obtained by tensoring two single-graviton states, each of which is a global descendant of the single-strand chiral state $\ket*{+}_1$, and then projecting onto the $S_2$-invariant part as appropriate to the $\lambda=\ydiagram{2}$ sector.  Its expression, $\tfrac{1}{2} (\phi^{2}_1 + \phi^{\ev{2}}_1)$, is obtained in the same way as in \eqref{eq:S_2_def}.  Here, for a function $f(x_1,x_2,\dots)$, the notation $f^{\ev{n}}$ denotes the function obtained by raising each argument to the $n$th power: $f^{\ev{n}}=f(x_1^n,x_2^n,\dots)$.
The global character expansion in the last column can be obtained using \eqref{eq.SU112CharProd} and \eqref{eq.SU112CharPow}.
Other lines in \eqref{eq:gravitons_sym_sector} are similarly obtained.

Some notational remarks are in order.  
We denote by ${\bf g}^{(h_0)}_{j,h}$ the gravity global multiplet corresponding to the global character $\phi^{(h_0)}_{j,h}$, and by $\ket*{g^{(h_0)}_{j,h}}$ the corresponding global primary.
We will label by $t ={\rm I, II, III},\dots$ the gravity towers ${\bf a}^{(t)}_{J,H}$ found in the procedure and denote by $\ket*{a^{(t)}_{J,H}}_{\YTscriptsize\ydiagram{2}}$ the corresponding affine primary. When confusion can arise, we put the sector label $\lambda=\YTnormalsize\ydiagram{2}$ on tower labels and write ``tower  I$_{\YTscriptsize\ydiagram{2}}$'', for example.
The global multiplets belonging to the tower ${\bf a}^{(t)}_{J,H}$ are denoted by ${\bf s}^{(t)}_{j,h}$, with possible additional indices, and the corresponding global characters and global primaries are denoted, respectively, by $\phi^{(t)}_{j,h}$  and $\ket*{s^{(t)}_{j,h}}$.  For short multiplets, both global and affine, we drop the subscript for the conformal weight ($h$ and $H$).

\bigskip

Now let us carry out the procedure.  For $N=2$, within the range \eqref{J_t_range} and with the ordering of charges \eqref{(j,h)_ordering}, the procedure is to search for affine primaries with charges in the order
\begin{align}
    (j,h)=(0,0),(\tfrac12,\tfrac12),(0,1),(0,2),\dots
    \label{eq:(j,h)_values_affine}
\end{align}
However, to verify the consistency of the procedure, we will examine the global primaries that appear at every possible value of charges $(j,h)$, not just the ones in \eqref{eq:(J,H)_ordering}, starting from $(0,0)$ and going up in $h$.  For simplicity, we will not present the explicit form of all relevant global primaries, which can be found in Appendix~\ref{app:expl_states}.
\begin{itemize}
    \item {\bf $(j,h)=(0,0)$:}\\
    The only graviton global primary at these charges in \eqref{eq:gravitons_sym_sector} is $\ket*{g^{(0)}_{0}}=\ket{-}_1^{[1]} \ket{-}_1^{[2]}$ corresponding to the global character $\phi^{(0)}_0=1$ in \eqref{eq:0}. This is clearly an affine primary and generates the first gravity tower, tower I, which is a short affine multiplet.
    Its affine primary is
    \begin{align} \label{eq:APrim_short_1}
        \ket*{a^{(\rm{I})}_0}_{\YTscriptsize\ydiagram{2}} = \ket*{g^{(0)}_{0}}= \ket*{-}_1^{[1]} \ket*{-}_1^{[2]}.
    \end{align}
    The affine character of tower I can be expanded in global characters as
    \begin{align}
        \Phibf^{(\rm{I})}_0 &= \left[ \phi^{(\rm{I})}_0 - \dot{\chi}_{\frac{1}{2}} \phi^{(\rm{I})}_{\frac{1}{2}} + 2 \phi^{(\rm{I})}_1 - \dot{\chi}_{\frac{1}{2}} \phi^{(\rm{I})}_{\frac{3}{2}} + \phi^{(\rm{I})}_2 \right] \notag\\
        &\qquad + \left[ \dot{\chi}_1 \phi^{(\rm{I})}_{0,1} - 2 \dot{\chi}_{\frac{1}{2}} \phi^{(\rm{I})}_{\frac{1}{2},\frac{3}{2}} + (1+\dot{\chi}_1 ) \phi^{(\rm{I})}_{1,2} + (4 + \chi_{\frac{1}{2}} \dot{\chi}_{\frac{3}{2}}) \phi^{(\rm{I})}_{0,2} + \cdots \right].\label{eq:tower1}
    \end{align}
    We put the label (I) on the affine and global characters to indicate that they belong to tower I\@.
    
    \item {\bf $(j,h)=\left( \frac{1}{2}, \frac{1}{2} \right)$:}\\
    The graviton global primaries at these charges are $\ket*{g^{(\frac{1}{2})}_{\frac{1}{2}}}^{\dot{A}}$ corresponding to the global character $-\dot{\chi}_{\frac{1}{2}} \phi^{(\frac{1}{2})}_{\frac{1}{2}}$ in \eqref{eq:1/2}, where $\dot{\chi}_{\half}$ corresponds to the doublet index $\Ad$ of this state and the minus sign means that this state is fermionic.\footnote{See Appendix \ref{app:expl_states} for the explicit expression for these global primaries as well as for other global primaries discussed below.} If we orthogonalize them against tower I, we obtain zero, indicating that there is no affine primary at these charges.  This is in fact obvious because $\ket*{g^{(\frac{1}{2})}_{\frac{1}{2}}}^{\dot{A}}$ coincide with global primaries inside tower I that we have already found:
    \begin{align}
        \bigket{g^{(\frac{1}{2})}_{\frac{1}{2}}}^{\dot{A}} = 
        \bigket{ s^{(\rm{I})}_{\frac{1}{2}} }^{\dot{A}},
    \end{align}
    where $\bigket{ s^{(\rm{I})}_{\frac{1}{2}} }^{\dot{A}}$ are singleton global primaries corresponding to $-\dot{\chi}_{\frac12}\phi^{\rm (I)}_{\frac12}$ in \eqref{eq:tower1}.
    Thus, there is no new tower at these charges.

    \item {\bf $(j,h)=( 1, 1 )$:}\\
    The graviton global primaries at these charges are $\ket*{g^{(1)}_1}$ corresponding to $\phi^{(1)}_{1}$ in \eqref{eq:1}
    and
    $\ket*{g^{(1')}_1}$ corresponding to $\phi^{(1')}_{1}$ in \eqref{eq:1'}. If we orthogonalize them against tower~I, we obtain zero, since they coincide with the global primaries inside tower I:
    \begin{align}
        \bigket{g^{(1)}_1} &= \bigket{ s^{(\rm{I})}_1 },\qquad
        \bigket{g^{(1')}_1} = \bigket{s'^{(\rm{I})}_1 },
    \end{align}
    where $\bigket{ s^{(\rm{I})}_1 }, \bigket{s'^{(\rm{I})}_1 }$ correspond to $2 \phi^{(\rm{I})}_1$ in \eqref{eq:tower1}.  
    Thus, there is no new tower.

    \item {\bf $(j,h)=( 0, 1)$:}\\
    The graviton global primaries at these charges are $\ket*{g^{(1)}_{0,1}}^a$ corresponding to $\dot{\chi}_1 \phi^{(1)}_{0,1}$ in \eqref{eq:1}. If we orthogonalize them against tower~I, namely against the global primary $\ket*{s^{(\rm{I})}_{0,1}}^a$ corresponding to $\dot{\chi}_1 \phi^{(\rm{I})}_{0,1}$ in \eqref{eq:tower1}, we find 
    \begin{align}
        \bigket{ g^{(1)}_{0,1} }^a_{\perp} &\equiv\bigket{ g^{(1)}_{0,1} }^a - \frac{1}{2} \bigket{ s^{(\rm{I})}_{0,1} }^a
        =-\frac{1}{2} \sigma_{\dot{A}\dot{B}}^a \psi^{-\dot{A}(\mathcal{A})}_{-\frac{1}{2}} \psi^{+\dot{B}(\mathcal{A)}}_{-\frac{1}{2}}  \ket{-}^{[1]}_1 \ket{-}^{[2]}_1 .
        \label{eq:g(1)0,1_perp}
    \end{align}
    Here and in what follows, we do not present states in a normalized form. For an operator $\cO$, we define the copy-antisymmetric combination by
    \begin{align}
        \mathcal{O}^{(\mathcal{A})} \equiv \mathcal{O}^{[1]} - \mathcal{O}^{[2]}.
    \end{align}
    We can show that \eqref{eq:g(1)0,1_perp} are affine primaries. Thus, we found a new long affine multiplet, tower II, with the affine primaries
    \begin{align} \label{eq:APrim_long_1}
        \big| a^{(\rm{II})}_{0,1} \big>^a_{\YTscriptsize\ydiagram{2}} = \big| g^{(1)}_{0,1} \big>^a_{\perp} = -\frac{1}{2} \sigma_{\dot{A}\dot{B}}^a \psi^{-\dot{A}(\mathcal{A})}_{-\frac{1}{2}} \psi^{+\dot{B}(\mathcal{A)}}_{-\frac{1}{2}}  \ket{-}^{[1]}_1 \ket{-}^{[2]}_1 .
    \end{align}
    Its affine character has the following global character expansion:
    \begin{align}
        \dot{\chi}_1 \Phibf{^{(\rm{II})}_{0,1}} = \dot{\chi}_1 \left( \phi^{(\rm{II})}_{0,1} - \dot{\chi}_{\frac{1}{2}} \phi^{(\rm{II})}_{\frac{1}{2},\frac{3}{2}} + \phi^{(\rm{II})}_{1,2} + (\chi_{\frac{1}{2}} \dot{\chi}_{\frac{1}{2}} + \dot{\chi}_1 )\phi^{(\rm{II})}_{0,2} + \cdots \right). \label{eq:tower2}
    \end{align}

    \item {\bf $(j,h)=\left( \frac{3}{2}, \frac{3}{2} \right)$:}\\
    The graviton global primaries at these charges are $\ket*{g^{(\frac{3}{2})}_{\frac{3}{2}}}^{\dot{A}}$ corresponding to $-\dot{\chi}_{\frac{1}{2}} \phi^{(\frac{3}{2})}_{\frac{3}{2}}$ in \eqref{eq:3/2}. Orthogonalizing them against towers I and II gives zero, since they coincide with a global primaries inside tower I:
    \begin{align}
        \bigket{ g^{(\frac{3}{2})}_{\frac{3}{2}} }^{\dot{A}} = \bigket{ s^{(\rm{I})}_{\frac{3}{2}} }^{\dot{A}},
    \end{align}
    where $\ket*{ s^{(\rm{I})}_{\frac{3}{2}} }^{\dot{A}}$ corresponds to
    $-\dot{\chi}_{\frac{1}{2}}\phi^{(\rm I)}_{\frac32}$ in \eqref{eq:tower1}.
    There is no new tower.

    \item {\bf $(j,h)=\left( \frac{1}{2}, \frac{3}{2} \right)$:}\\
    The graviton global primaries at these charges are $\ket*{g^{(\frac{3}{2})}_{\frac{1}{2},\frac{3}{2}}}^{\dot{A}}$ corresponding to $-\dot{\chi}_{\frac{1}{2}} \phi^{(\frac{3}{2})}_{\frac{1}{2},\frac{3}{2}}$ in \eqref{eq:3/2}. Orthogonalizing them against towers I and II gives zero:
    \begin{align}
        \bigket{ g^{(\frac{3}{2})}_{\frac{1}{2},\frac{3}{2}} }^{\dot{A}}_{\perp} &\equiv 
        \bigket{ g^{(\frac{3}{2})}_{\frac{1}{2},\frac{3}{2}} }^{\dot{A}}
        +\frac{3}{4} \bigket{ s^{(\rm{I})}_{\frac{1}{2},\frac{3}{2}} }^{\dot{A}} 
        -\frac{3}{4} \bigket{ s'^{(\rm{I})}_{\frac{1}{2},\frac{3}{2}}}^{\dot{A}} 
        +\frac{3}{4} {\left( \sigma^{a} \right)^{\dot{A}}}_{\dot{B}}
        \bigket{ s^{(\rm{II})}_{\frac{1}{2},\frac{3}{2}} }^{\dot{B},a}
        =0,
    \end{align}
where
$\ket*{s^{(\rm{I})}_{\frac{1}{2},\frac{3}{2}}}^{\dot{A}}$ and $\ket*{s'^{(\rm{I})}_{\frac{1}{2},\frac{3}{2}}}^{\dot{A}}$ correspond to $-2 \dot{\chi}_{\frac{1}{2}} \phi_{\frac{1}{2},\frac{3}{2}}^{(\rm{I})}$ in \eqref{eq:tower1}, while $\ket*{s^{(\rm{II})}_{\frac{1}{2},\frac{3}{2}}}^{\dot{A},a}$ corresponds to  $-\dot{\chi}_1  \dot{\chi}_{\frac{1}{2}} \phi_{\frac{1}{2},\frac{3}{2}}^{(\rm{II})}$ in \eqref{eq:tower2}.
Thus, there is no new tower at these charges.  We have written out explicitly how the combination vanishes, which can be regarded as a consistency check of the procedure: there must be no affine primary at these charges because for $N=2$ affine long primaries must have $j=0$.  This is also an example of how keeping track of $SU(2)_2$ representations helps organize states.

    \item {\bf $(j,h)=(2, 2)$:}\\
    The graviton global primary at these charges is $\ket*{g^{(2)}_{2}}$ corresponding to $\phi^{(2)}_{2}$ in \eqref{eq:2}. Orthogonalizing them against towers I and II gives zero, since it coincides with a global primary inside tower I:
    \begin{align}
        \bigket{ g^{(2)}_{2} } = \bigket{ s^{(\rm{I})}_{2} },
    \end{align}
    where $\ket*{ s^{(\rm{I})}_{2} }$ corresponds to $\phi^{(\rm{I})}_2$ in \eqref{eq:tower1}.
    There is no new tower.

    \item {\bf $(j,h)=( 1, 2)$:}\\
    There is no graviton global primary at these charges.

    \item {\bf $(j,h)=( 0, 2)$:}\\
    The graviton global primaries at these charges are $\ket*{g^{(1)}_{0,2}}$ corresponding to  $\phi^{(1)}_{0,2}$ in \eqref{eq:1} and $\ket*{g^{(2)}_{0,2}}$ corresponding to $\phi^{(2)}_{0,2}$ in \eqref{eq:2}. Orthogonalizing them against towers I and II gives only one independent state, which can be taken to be\footnote{See Appendix \ref{app:expl_states} for details.}
    \begin{align}
        \bigket{ g^{(1)}_{0,2}}_{\perp}
        &\equiv \bigket{ g^{(1)}_{0,2} }
        -\frac{11}{36} \bigket{ s''^{(\rm{I})}_{0,2} }
        -\frac{1}{5} \bigket{ s'''^{(\rm{I})}_{0,2} },
       % \bigket{ g^{(2)}_{0,2} }_{\perp} &=0,
    \end{align}
where $\ket*{ s''^{(\rm{I})}_{0,2} }$ and $\ket*{ s'''^{(\rm{I})}_{0,2} }$
are two of the four global primaries corresponding to $4\phi_{0,2}^{(\rm{I})}$ in \eqref{eq:tower1}.
Note that other $\phi_{0,2}$ terms in \eqref{eq:tower1} and \eqref{eq:tower2} are irrelevant, having different (non-trivial) $SU(2)_1\times SU(2)_2$ representations.
    We can show that
    $\ket*{g^{(1)}_{0,2}}_{\perp}$ is an affine primary. Thus, we found a new long affine multiplet, tower III, whose affine primary is:
    \begin{align} \label{eq:APrim_long_2}
        \big| a^{(\rm{III})}_{0,2} \big>_{\YTscriptsize\ydiagram{2}} &=\big| g^{(1)}_{0,2} \big>_{\perp} \notag\\
        &=\left[ \frac{1}{40} \epsilon_{\alpha \gamma} \epsilon_{\beta \delta} \epsilon_{\dot{A}\dot{B}} \epsilon_{\dot{C}\dot{D}} \psi^{\alpha\dot{A}(\mathcal{A})}_{-\frac{1}{2}} \psi^{\beta \dot{B}(\mathcal{A})}_{-\frac{1}{2}} \psi^{\gamma \dot{C}(\mathcal{A})}_{-\frac{1}{2}} \psi^{\delta \dot{D}(\mathcal{A})}_{-\frac{1}{2}} \right. \notag\\
        &\left. 
        \qquad-\frac{3}{10} \epsilon_{\alpha \beta} \epsilon_{\dot{A}\dot{B}} \psi^{\alpha\dot{A}(\mathcal{A})}_{-\frac{3}{2}} \psi^{\beta \dot{B}(\mathcal{A})}_{-\frac{1}{2}} - \frac{3}{20} \epsilon_{\dot{A}\dot{B}}\epsilon_{AB} \alpha^{\dot{A}A(\mathcal{A})}_{-1} \alpha^{\dot{B}B(\mathcal{A})}_{-1} \right] \ket{-}_1^{[1]} \ket{-}_1^{[2]}.
    \end{align}
    
\end{itemize}
Therefore, up to $h=2$, we found that the inclusion of singletons leads to three affine towers:   I$_{\YTscriptsize\ydiagram{2}}$,
II$_{\ydiagram{2}}$ and 
III$_{\ydiagram{2}}$.

\YTnormalsize
\subsubsection{Twisted sector ($\lambda=\ydiagram{1}\mkern1.5mu$)} \label{sssec:N=(2)}

In the twisted sector, the chiral states and the partition sum of the graviton states built on them are as follows:
\begin{subequations}
\label{eq:gravitons_twisted_sector}
    \begin{IEEEeqnarray}{cclll}
        \text{chiral states} &\qquad&  \text{graviton partition sum} &\qquad& \text{global character expansion}\notag\\
    \noalign{\vskip-2ex}
    \IEEEeqnarraymulticol{5}{c}{\makebox[13cm][l]{\rule{13cm}{0.4pt}}}\notag\\
    \noalign{\vskip-0.5ex}
    \ket*{-}_2 && \phi_\frac{1}{2}&& \phi^{(\frac{1}{2})}_\frac{1}{2} \label{eq:1/2twist}\\
    \ket*{\dot{A}}_2 && -\dot{\chi}_\frac{1}{2}\phi_1&&  -\dot{\chi}_\frac{1}{2}\phi^{(1)}_1\label{eq:1twist}\\
    \ket*{+}_2 && \phi_\frac{3}{2}&& \phi^{(\frac{3}{2})}_\frac{3}{2}\label{eq:3/2twist}
    \end{IEEEeqnarray}
\end{subequations}
In the global character expansion, we are using the same labeling scheme for global primaries that we used for the symmetric sector.  We see that, in this sector, there are only three graviton global primaries at $(j,h)=\left( \frac{1}{2}, \frac{1}{2} \right)$, $\left( 1,1 \right)$, $\left( \frac{3}{2},\frac{3}{2} \right)$, which we examine in turn.

\begin{itemize}
    \item {\bf $(j,h)=\left( \frac{1}{2}, \frac{1}{2} \right)$:}\\
    The only graviton global primary at these charges is $\ket*{g^{(0)}_{\frac12}}=\ket*{-}_2$ in (\ref{eq:1/2twist}). This is an affine primary and generates the first gravity tower in this sector, which we call tower~I\@. Note that the label I was also used in the $\lambda=\YTnormalsize\ydiagram{2}$ sector, where it refers to a different tower. When confusions can arise, we will denote tower I in the $\lambda=\ydiagram{1}$ sector by I$_{\YTscriptsize\ydiagram{1}}$.
        Tower I is a short affine multiplet and has the affine primary
    \begin{align} \label{eq:APrim_short_2}
    \YTscriptsize
        \ket*{a_{\frac{1}{2}}^{(\rm I)}}_{\ydiagram{1}}=\ket*{g^{(0)}_{\frac12}}=\ket*{-}_2.
    \end{align}
    Its affine character can be expanded in global characters as
    \begin{multline}\label{eq:towerItilde}
        \Phibf^{(\rm I)}_\frac{1}{2}=\left( \phi^{(\rm I)}_\frac{1}{2}-\dot{\chi}_\frac{1}{2}\phi^{(\rm I)}_1+\phi^{(\rm I)}_\frac{3}{2} \right)\\
        +\left( -\dot{\chi}_\frac{1}{2}\phi^{(\rm I)}_{0,1}+(2+\dot{\chi}_1)\phi^{(\rm I)}_{\frac{1}{2},\frac{3}{2}}-(3\dot{\chi}_\frac{1}{2}+\chi_\frac{1}{2} \dot{\chi}_1)\phi^{(\rm I)}_{0,2}-2\dot{\chi}_\frac{1}{2}\phi^{(\rm I)}_{1,2}+\cdots \right).
    \end{multline}

    \item {\bf $(j,h)=(1,1)$:}\\
    The graviton global primaries at these charges are $\ket*{g_1^{(1)}}^{\dot{A}}$ corresponding to $-\dot{\chi}_\frac{1}{2} \phi^{(\rm I)}_1$ in (\ref{eq:1twist}). Orthogonalizing them against tower I gives zero, since they coincide with singleton global primaries inside tower I:
    \begin{align}
        \ket*{g_1^{(1)}}^{\dot{A}}=\ket*{s_1^{(\rm I)}}^{\dot{A}},
    \end{align}
 where $\ket*{s_1^{(\rm I)}}^{\dot{A}}$ correspond to $-\dot{\chi}_{\frac12}\phi^{(\rm I)}_1$ in \eqref{eq:towerItilde}.  There is no new tower at these charges.

    \item {\bf $(j,h)=\left( \frac{3}{2}, \frac{3}{2} \right)$:}\\
    The graviton global primary at these charges is $\ket*{g_\frac{3}{2}^{(1)}}$ corresponding to $\phi^{(\rm I)}_\frac{3}{2}$ in (\ref{eq:3/2twist}).  Orthogonalizing it against tower I gives zero, since it coincides with a global primary inside tower I:
    \begin{align}
        \ket*{g_\frac{3}{2}^{(1)}}=\ket*{s_\frac{3}{2}^{(\rm I)}},
    \end{align}
where $\ket*{s_\frac{3}{2}^{(\rm I)}}$ corresponds to $\phi^{\rm(I)}_{\frac32}$ in  (\ref{eq:towerItilde}). We find no new tower.
\end{itemize}

Therefore, in the twisted sector, there is only one affine tower: I$_{\ydiagram{1}}$.

\YTnormalsize
\subsubsection{Untwisted anti-symmetric sector ($\lambda=\ydiagram{1,1}\mkern1.5mu$)} \label{sssec:N=(1,1)_asym}

In the anti-symmetric sector, the chiral states, the partition sum for gravitons built on them, and its global character expansion are as follows: 
% \begin{subequations}
% \begin{align}
%     \ket*{-}_1^{[1]} \ket*{\dot{A}}_1^{[2]} &: -\dot{\chi}_{\frac{1}{2}} \phi^{(\frac{1}{2})}_{\frac{1}{2}},\label{eq:1/2anti}\\
%         \ket*{-}_1^{[1]} \ket*{+}_1^{[2]} &: \phi^{(1')}_{1},\label{eq:1'anti}\\
%         \ket*{\dot{A}}_1^{[1]} \ket*{\dot{B}}_1^{[2]} &: \frac{1}{2} \left[ (-\dot{\chi}_\frac{1}{2} \phi_{\frac{1}{2}} )^2 - (-\dot{\chi}^{[2]}_\frac{1}{2} \phi^{[2]}_{\frac{1}{2}} ) \right] = \dot{\chi}_1\phi^{(1)}_{1} + (\dot{\chi}_0 \phi^{(1)}_{0,1} + \dot{\chi}_1\phi^{(1)}_{0,2} + \cdots ),\label{eq:1anti}\\
%         \ket*{+}_1^{[1]} \ket*{\dot{A}}_1^{[2]} &: \phi_1 (-\dot{\chi}_{\frac{1}{2}} \phi_{\frac{1}{2}}) = - \dot{\chi}_{\frac{1}{2}} \phi^{(\frac{3}{2})}_{\frac{3}{2}} - ( \dot{\chi}_{\frac{1}{2}} \phi^{(\frac{3}{2})}_{\frac{1}{2},\frac{3}{2}} + \dot{\chi}_{\frac{1}{2}} \phi^{(\frac{3}{2})}_{\frac{1}{2},\frac{5}{2}} + \cdots ),\label{eq:3/2anti}\\
%         \ket*{+}_1^{[1]} \ket*{+}_1^{[2]} &: \frac{1}{2} (\phi^{2}_1 - \phi^{[2]}_1 ) = 0 + (\phi^{(2)}_{1,2} +\cdots),\label{eq:2anti}
% \end{align}
% \end{subequations}
\begin{subequations}
\label{eq:gravitons_antisym_sector}
    \begin{IEEEeqnarray}{cclll}
        \text{chiral states} &\quad&  \text{graviton partition sum} &\qquad& \text{global character expansion}\notag\\
    \noalign{\vskip-2ex}
    \IEEEeqnarraymulticol{5}{c}{\makebox[15.0cm][l]{\rule{14.5cm}{0.4pt}}}\notag\\
    \noalign{\vskip-0.5ex}
    \ket*{-}_1 \ket*{\dot{A}}_1 && -\dot{\chi}_{\frac{1}{2}} \phi_{\frac{1}{2}} && -\dot{\chi}_{\frac{1}{2}} \phi^{(\frac{1}{2})}_{\frac{1}{2}}\label{eq:1/2anti}\\
    \ket*{\dot{A}}_1 \ket*{\dot{B}}_1 && \tfrac{1}{2} [ (-\dot{\chi}_\frac{1}{2} \phi_{\frac{1}{2}} )^2 - (-\dot{\chi}^{\ev{2}}_\frac{1}{2} \phi^{\ev{2}}_{\frac{1}{2}} ) ] && \dot{\chi}_1\phi^{(1)}_{1} + ( \phi^{(1)}_{0,1} + \dot{\chi}_1\phi^{(1)}_{0,2} + \cdots ) \label{eq:1anti}\\
    \ket*{-}_1 \ket*{+}_1 && \phi_{1}&& \phi^{(1')}_{1}\label{eq:1'anti}\\
    \ket*{+}_1 \ket*{\dot{A}}_1 && \phi_1 (-\dot{\chi}_{\frac{1}{2}} \phi_{\frac{1}{2}}) && - \dot{\chi}_{\frac{1}{2}} \phi^{(\frac{3}{2})}_{\frac{3}{2}} - ( \dot{\chi}_{\frac{1}{2}} \phi^{(\frac{3}{2})}_{\frac{1}{2},\frac{3}{2}} + \dot{\chi}_{\frac{1}{2}} \phi^{(\frac{3}{2})}_{\frac{1}{2},\frac{5}{2}} + \cdots ) \label{eq:3/2anti}\\
    \ket*{+}_1 \ket*{+}_1 && \tfrac{1}{2} (\phi^{2}_1 - \phi^{\ev{2}}_1 ) && 0 + (\phi^{(2)}_{1,2} +\cdots) \label{eq:2anti}
    \end{IEEEeqnarray}
\end{subequations}

For example, in \eqref{eq:2anti}, we have written the partition sum for two-graviton states obtained by tensoring two single-graviton states each built on $\ket*{+}_1$, and then projecting onto the $S_2$-anti-symmetric part as appropriate to the $\lambda=\YTnormalsize\ydiagram{1,1}$ sector.  Its expression,  $\tfrac{1}{2} (\phi^{2}_1 - \phi^{\ev{2}}_1 )$, is found just as in \eqref{eq:S_1,1_def}.  Its global primary expansion does not contain $\phi_2$ because the would-be bottom state $\ket*{+}_1^{[1]}\ket*{+}_1^{[2]}$ vanishes upon copy-antisymmetrization. Likewise, we do not have $\ket{-}_1\ket{-}_1$ in the list \eqref{eq:gravitons_sym_sector}, because $\ket*{-}_1$ has no global descendant and the would-be bottom component $\ket*{-}_1\ket*{-}_1$ vanishes upon copy-antisymmetrization.

\bigskip
Now let us carry out the procedure starting from $(j,h)=\left(\frac{1}{2},\frac{1}{2}\right)$.

\begin{itemize}
    \item {\bf $(j,h)=\left(\frac{1}{2},\frac{1}{2}\right)$:} \\
    The only graviton global primaries at these charges are 
    \begin{align}
        \ket*{g_{\frac{1}{2}}^{(\frac{1}{2})}}^{\Ad}= \ket*{-}_1^{[1]} \ket*{\dot{A}}_1^{[2]} - \ket*{\dot{A}}_1^{[1]} \ket*{-}_1^{[2]},
    \end{align}
    corresponding to $-\dot{\chi}_{\frac{1}{2}} \phi^{(\frac{1}{2})}_{\frac{1}{2}}$ in \eqref{eq:1/2anti}. These are affine primaries and generate short affine multiplets, which we call tower I (or tower I$_{\YTscriptsize\ydiagram{1,1}}$).
    The affine primaries in tower I are
    \begin{align} \label{eq:APrim_short_3}
        \ket*{a^{(\rm{I})}_{\frac{1}{2}}}^{\dot{A}}_{\ydiagram{1,1}} = 
        \ket*{g_{\frac{1}{2}}^{(\frac{1}{2})}}^{\Ad} =\ket*{-}_1^{[1]} \ket*{\dot{A}}_1^{[2]} - \ket*{\dot{A}}_1^{[1]} \ket*{-}_1^{[2]}\ .
    \end{align}
        The affine character of tower I has the global character expansion:
    \begin{align}
        -\dot{\chi}_{\frac{1}{2}} \Phibf{^{(\rm{I})}_{\frac{1}{2}}} 
        % &= -\dot{\chi}_{\frac{1}{2}} \left[ \left(\phi^{(\rm{I})}_{\frac{1}{2}} - \dot{\chi}_{\frac{1}{2}} \phi^{(\rm{I})}_{1} + \phi^{(\rm{I})}_{\frac{3}{2}} \right)\right.\\
        % \nonumber
        % &\left.\ \ \ \ \ \ \ \ \ \ \ +\left( -\dot{\chi}_{\frac{1}{2}}\phi^{(\rm{I})}_{0,1}+(2+\dot{\chi}_{1})\phi^{(\rm{I})}_{\frac{1}{2},\frac{3}{2}}-2\dot{\chi}_{\frac{1}{2}}\phi^{(\rm{I})}_{1,2}-(3\dot{\chi}_{\frac{1}{2}}+\chi_{\frac{1}{2}}\dot{\chi}_{1})\phi^{(\rm{I})}_{0,2} + \cdots \right) \right]\\
        &=-\dot{\chi}_{\frac{1}{2}}\phi^{(\rm{I})}_{\frac{1}{2}}+(\dot{\chi}_0+\dot{\chi}_{1})\phi^{(\rm{I})}_1-\dot{\chi}_{\frac{1}{2}}\phi^{(\rm{I})}_{\frac{3}{2}} + (\dot{\chi}_0+\dot{\chi}_1)\phi^{(\rm{I})}_{0,1} - (3\dot{\chi}_{\frac12} + \dot{\chi}_{\frac32})\phi^{(\rm{I})}_{\frac12,\frac32} \nonumber\\
        &\quad + 2(\dot{\chi}_{0} + \dot{\chi}_{1})\phi^{(\rm{I})}_{1,2} + \big(3\dot{\chi}_{0} + 3\dot{\chi}_1 + \chi_{\frac12}(\dot{\chi}_{\frac12}+\dot{\chi}_{\frac32})\big)\phi^{(\rm{I})}_{0,2} +\cdots \label{eq:tower1'}
    \end{align}

    \item {\bf $(j,h)=\left(1,1\right)$:}\\
    The only graviton global primaries are 
    $\ket*{g_1^{(1)}}^a$ corresponding to $\dot{\chi}_1\phi^{(1)}_1$ in \eqref{eq:1anti} and $\ket*{g_1^{(1')}}$ corresponding to $\phi^{(1')}_1$ in \eqref{eq:1'anti}. Orthogonalizing them against tower I gives zero, since they coincide with global primaries inside tower I:
    \begin{align}
        \bigket{ g^{(1)}_1 }^a = \bigket{ s^{(\rm{I})}_1 }^a,
        \qquad
        \bigket{ g^{(1')}_1 } &= \bigket{ s'^{(\rm{I})}_1 },
    \end{align}
    where $\ket*{ s^{(\rm{I})}_1 }^a$ and $\ket*{ s'^{(\rm{I})}_1 }$ correspond to $\dot{\chi}_{1}\phi^{(\rm{I})}_1$ and $\phi^{(\rm{I})}_1$ in \eqref{eq:tower1'}, respectively. 
    Thus, there is no new tower at these charges.

    \item {\bf $(j,h)=\left(0,1\right)$:}\\
    The only graviton global primary is $\ket*{g_{0,1}^{(1)}}$ corresponding to $\phi^{(1)}_{0,1}$ in \eqref{eq:1anti}. Orthogonalizing it against tower I gives zero, since it coincides with a global primary inside tower~I:
    \begin{align}
        \bigket{ g^{(1)}_{0,1} } = \bigket{ s^{(\rm{I})}_{0,1} },
    \end{align}
where $\ket*{ s^{(\rm{I})}_{0,1} }$ corresponds to $\phi^{(\rm{I})}_{0,1}$ in \eqref{eq:tower1'}.    
    Thus, there is no new tower at these charges.

    \item {\bf $(j,h)=\left(\frac{3}{2},\frac{3}{2}\right)$:} \\
    The only graviton global primaries are $\ket*{g_{\frac{3}{2}}^{(\frac{3}{2})}}^{\dot{A}}$ corresponding to $-\dot{\chi}_{\frac{1}{2}}\phi^{(\frac{3}{2})}_{\frac{3}{2}}$ in \eqref{eq:3/2anti}. Orthogonalizing them against tower I gives zero, since they coincide with global primaries inside tower I:
    \begin{align}
        \bigket{ g^{(\frac{3}{2})}_{\frac{3}{2}} }^{\Ad}
        = \bigket{ s^{(\rm{I})}_{\frac{3}{2}} }^{\Ad},
    \end{align}
    where $\ket*{ s^{(\rm{I})}_{\frac{3}{2}} }$ corresponds to $-\dot{\chi}_{\frac{1}{2}}\phi^{(\rm{I})}_{\frac{3}{2}}$ in \eqref{eq:tower1'}.
        There is no new tower.

    \item {\bf $(j,h)=\left(\frac{1}{2},\frac{3}{2}\right)$:} \\
    The only graviton global primaries are $\ket*{g_{\frac{1}{2},\frac{3}{2}}^{(\frac{3}{2})}}^{\dot{A}}$ corresponding to $-\dot{\chi}_{\frac{1}{2}}\phi^{(\frac{3}{2})}_{\frac{1}{2},\frac{3}{2}}$ in \eqref{eq:3/2anti}. Orthogonalizing them against tower I gives zero:
    \begin{align}
        \big| g^{(\frac{3}{2})}_{\frac{1}{2},\frac{3}{2}} \big>^{\dot{A}}_{\perp}
        =
        \big| g^{(\frac{3}{2})}_{\frac{1}{2},\frac{3}{2}} \big>^{\dot{A}}
        +\frac{3}{4}\big| s_{\frac{1}{2},\frac{3}{2}}^{(\rm{I})}\big>^{\dot{A}}
        +\frac{1}{2}\big| s_{\frac{1}{2},\frac{3}{2}}'^{(\rm{I})}\big>^{\dot{A}}
        +\frac{3}{8}\big| s_{\frac{1}{2},\frac{3}{2}}''^{(\rm{I})}\big>^{\dot{A}}
        =0,
    \end{align}
    where $\ket*{s_{\frac{1}{2},\frac{3}{2}}^{(\rm{I})}}^{\dot{A}}$, $\ket*{ s_{\frac{1}{2},\frac{3}{2}}'^{(\rm{I})}}^{\dot{A}}$ and $\ket*{ s_{\frac{1}{2},\frac{3}{2}}''^{(\rm{I})}}^{\dot{A}}$ correspond to  $-3\dot{\chi}_{\frac{1}{2}}\phi^{(\rm{I})}_{\frac{1}{2},\frac{3}{2}}$ in \eqref{eq:tower1'}.
We find no new tower.

    \item {\bf $(j,h)=\left(1,2\right)$:} \\
    The only graviton global primary is $\ket*{g_{1,2}^{(2)}}$ corresponding to $\phi^{(2)}_{1,2}$ in \eqref{eq:2anti}. Orthogonalizing it against tower I gives zero:
    \begin{align}
        \big| g^{(2)}_{1,2}\big>_\perp=\big| g^{(2)}_{1,2}\big>+\big| s^{(\rm{I})}_{1,2}\big> =0,
    \end{align}
    where $\big| s^{(\rm{I})}_{1,2}\big>$ is one of the two singleton global primaries that corresponds to $2\dot{\chi}_{0}\phi^{(\rm{I})}_{1,2}$ in \eqref{eq:tower1'}. 
    There is no new tower.

    \item {\bf $(j,h)=\left(0,2\right)$:} \\
    The only graviton global primaries are $\ket*{g_{0,2}^{(1)}}^a$ corresponding to $\dot{\chi}_1\phi^{(1)}_{0,2}$ in \eqref{eq:1anti}. At these charges, there are three $SU(2)_2$-triplet global primaries within tower $\rm{I}$, corresponding to $3\dot{\chi}_{1}\phi^{(\rm{I})}_{0,2}$ in \eqref{eq:tower1'}. We can orthogonalize the supergraviton states against them to get zero:
    \begin{align}\label{eq:0,2antiorthogonal}
        \bigket{ g^{(1)}_{0,2}} ^a_\perp
        =
        \bigket{ g^{(1)}_{0,2}}^a
        -\frac{3}{5}\bigket{ s^{(\rm{I})}_{0,2}}^a
        +\frac{3}{8}\bigket{ s'^{(\rm{I})}_{0,2}}^a
        -\bigket{ s''^{(\rm{I})}_{0,2}}^a
        =0.
    \end{align}
    Thus there is no new tower at these charges.

\end{itemize}

Therefore, in the anti-symmetric sector, there is only one affine tower: I$_{\ydiagram{1,1}}$.

\subsubsection{Summary}

The spectrum of graviton affine towers that we found up to level $h=2$ can be summarized in the following left-moving supergravity BPS partition functions:
\begin{subequations}
\label{eq:sg_PF}
\begin{align}
    S_{\ydiagram{2}}^{\rm gravity}
    &=    \Phibf_0^{\rm (I)}     + \dot{\chi}_1\Phibf_{0,1}^{\rm (II)}    +    \Phibf_{0,2}^{\rm (III)}+\cdots,\label{eq:sg_PF_2}
\\
    S_{\ydiagram{1}}^{\rm gravity}
    &=    \Phibf_{\frac12}^{\rm (I)}     + \cdots,\label{eq:sg_PF_1}
\\
    S_{\ydiagram{1,1}}^{\rm gravity}
    &=   -\dot{\chi}_{\frac12} \Phibf_{\frac12}^{\rm (I)}     +\cdots,\label{eq:sg_PF_1,1}
\end{align}
\end{subequations}
The expressions for affine primaries are found in \eqref{eq:APrim_short_1}, \eqref{eq:APrim_long_1}, and \eqref{eq:APrim_long_2} for towers I$_{\ydiagram{2}}$, II$_{\ydiagram{2}}$ and III$_{\ydiagram{2}}$; in \eqref{eq:APrim_short_2} for tower I$_{\ydiagram{1}}$; and in \eqref{eq:APrim_short_3} for tower I$_{\ydiagram{1,1}}$.

\subsection{Comparison with the CFT spectrum} \label{ssec:Comparison}

Having constructed explicitly the affine primaries up to level 2, let us now compare the gravity-sector spectrum 
summarized in the partition functions \eqref{eq:sg_PF}
with the full CFT BPS spectrum using the partition functions \eqref{eq:S_CFT_exp_refined}.

Firstly, all short affine multiplets in the CFT are accounted for within the gravity sector, namely by the states~\eqref{eq:APrim_short_1}, \eqref{eq:APrim_short_2} and \eqref{eq:APrim_short_3} for the $\lam=\YTnormalsize\ydiagram{2}$, $ \ydiagram{1,1}$ and $\ydiagram{1}$ symmetry sectors, respectively.   This is evident from the agreement of the short affine characters in \eqref{eq:sg_PF} and those in \eqref{eq:S_CFT_exp_refined}, and is unsurprising since the affine primary states of short multiplets are chiral primaries ($1/2$-BPS) and are hence protected. 

Let us turn to the spectra of long affine multiplets. In the $\lam=\YTnormalsize\ydiagram{2}$ sector, all long affine characters in the CFT that appear in the partition function \eqref{eq:S_CFT_2_exp_refined} (up to level two that we have worked to) are captured by the affine primaries constructed within the gravity sector: \textit{i.e.}, at level $h=1$ by the affine primary \eqref{eq:APrim_long_1} and at $h=2$ by the affine primary \eqref{eq:APrim_long_2}.
In contrast, for neither the $\lam=\ydiagram{1}\,$ nor the $\lam=\ydiagram{1,1}\,$ sectors, did we find within the gravity sector any affine primaries of long multiplets that appear in the CFT partition function \eqref{eq:S_CFT_1,1_exp_refined} and \eqref{eq:S_CFT_1_exp_refined}. 
Specifically, in the $\lambda=\YTnormalsize\ydiagram{1}$ sector, at level $h=1$, the affine primaries in the CFT are\footnote{Here we are using $\ket{\mathfrak{a}}$ to distinguish these CFT affine primary states from those constructed in Section~\ref{ssec:L_results} (denoted by $\ket{a}$ there), contained within the gravity sector of the theory. }
\begin{equation} \label{eq:APrim_long_4}
\YTscriptsize
    \ket{\mathfrak{a}_{0,1}}^{A,a}_{\ydiagram{1}} = \sigma^{a}_{\Ad\Bd} \psi^{-\Ad}_{0} \alpha_{-\half}^{\Bd A} \ket{-}_2^{[1]} \ ,
\end{equation}
which account for the term $-\chi_{\frac12} \dot{\chi}_{1} \mathbf{\Phi}_{0,1}$ in the CFT partition function \eqref{eq:S_CFT_1_exp_refined} but does not appear in the gravity-sector partition function \eqref{eq:sg_PF_1}.  At level $h=2$, the CFT affine primaries are 
\begin{subequations} \label{eq:APrim_long_5-7}
\begin{align} \label{eq:APrim_long_5}
    \ket{\mathfrak{a}_{0,2}}^{A} _{\ydiagram{1}} &=  \Big[\ep_{\Ad\Bd} \psi^{-\Ad}_{0} \alpha_{-\frac32}^{\Bd A} - 3\ep_{\Ad\Bd} \psi^{-\Ad}_{-1} \alpha_{-\frac12}^{\Bd A} - \frac34\ep_{\Ad\Dd}\ep_{\Bd\Cd} \psi^{+\Ad}_{-1}\psi^{-\Bd}_{0}\psi^{-\Cd}_{0}\alpha^{\Dd A}_{-\frac12} \nonumber\\
    &\hspace{25ex} + \frac12\ep_{BC}\ep_{\Ad\Bd}\ep_{\Cd\Dd}\alpha_{-\frac12}^{\Ad B}\alpha_{-\frac12}^{\Bd C} \psi^{-\Cd}_{0}\alpha_{-\frac12}^{\Dd A} \Big] \ket{-}_2^{[1]} \ ,\\[1ex]
    \ket{\mathfrak{a}'_{0,2}}^{ABC,\Ad\Bd\Cd\Dd}_{\ydiagram{1}} 
    &= 
    S^{ABC}_{EFG}\,
    S^{\Ad\Bd\Cd\Dd}_{\dot{E}\dot{F}\dot{G}\dot{H}} \,
    \alpha_{-\frac12}^{\dot{E} E} \alpha_{-\frac12}^{\dot{F} F} \alpha_{-\frac12}^{\dot{G} G} \psi^{-\dot{H}}_{0} \ket{-}_2^{[1]} \ , \label{eq:APrim_long_6}\\[1ex]
    \ket{\mathfrak{a}''_{0,2}}^{A,a} _{\ydiagram{1}}&= \sigma^a_{\Ad\Bd}\Big[ \psi^{-\Ad}_{-1} \alpha_{-\frac12}^{\Bd A} + \frac13 \psi^{-\Ad}_{0} \alpha_{-\frac32}^{\Bd A} - \frac12 \ep_{BC}\ep_{\Cd\Dd} \psi^{-\Ad}_{0}\alpha_{-\frac12}^{\Bd A} \alpha_{-\frac12}^{\Cd B} \alpha_{-\frac12}^{\Dd C} \nonumber\\
    &\hspace{25ex}
    - \frac12 \ep_{\Cd\Dd} \psi^{-\Cd}_{0} \psi^{+\Dd}_{-1} \psi^{-\Ad}_{0} \alpha_{-\frac12}^{\Bd A} \Big] \ket{-}_2^{[1]} \ , \label{eq:APrim_long_7}
\end{align}
\end{subequations}
where we defined the symmetrizer for tensors by
\begin{align}
    S^{a_1\dots a_k}_{b_1\dots b_k}
    \equiv {1\over k!}\sum_{\sigma\in S_k}
    \delta^{a_1}_{b_{\sigma(1)}}\!\cdots\, \delta^{a_k}_{b_{\sigma(k)}} \ .
\end{align}
In \eqref{eq:APrim_long_6}, $S^{ABC}_{EFG}$ and $S^{\Ad\Bd\Cd\Dd}_{\dot{E}\dot{F}\dot{G}\dot{H}}$ act as the projectors onto the spin-$\frac32$ representation of $SU(2)_1$ and the spin-1 representation of $SU(2)_2$, respectively.
The affine primaries \eqref{eq:APrim_long_5-7} respectively account for the terms 
$-\chi_{\frac12}\mathbf{\Phi}_{0,2}$, $-\chi_{\frac32} \dot{\chi}_{2}\mathbf{\Phi}_{0,2}$  and $-\chi_{\frac12} \dot{\chi}_{1}\mathbf{\Phi}_{0,2}$ in \eqref{eq:S_CFT_1_exp_refined}, but none of them appears in \eqref{eq:sg_PF_1}.
Likewise, in the $\lambda=\YTnormalsize\ydiagram{1,1}$ sector, 
the affine primaries in the CFT at level $h=2$, which is not in the gravity sector, can be shown to be
\begin{equation} \label{eq:APrim_long_3}
    \ket{\mathfrak{a}_{0,2}}^{A,\Ad\Bd\Cd}_{\YTscriptsize\ydiagram{1,1}} = \ep_{\alpha\beta} \,S^{\Ad\Bd\Cd}_{\dot{D}\dot{E}\dot{F}}\, \psi_{-\frac12}^{\alpha\dot{D}(\cA)} \psi_{-\frac12}^{\beta\dot{E}(\cA)} \alpha_{-1}^{\dot{F} A(\cA)} \ket{-}_1^{[1]}\ket{-}_1^{[2]} \ .
\end{equation}
The states \eqref{eq:APrim_long_3} fall into a $(\mathbf{2},\mathbf{4})$ representation of $SU(2)_1\times SU(2)_2$ and account for the term $\chi_{\frac12} \dot{\chi}_{\frac32} \mathbf{\Phi}_{0,2}$ in \eqref{eq:S_CFT_1,1_exp_refined}. 
The states \eqref{eq:APrim_long_4}--\eqref{eq:APrim_long_3} are stringy states, involving non-total affine generators or $\alpha_{-r}$ with fractional $r$, which explains why our procedure did not produce those states.

In fact, the above long multiplets---specifically, all those in the $\lambda=\YTnormalsize\ydiagram{2}$ sector which are realized in the gravity sector and some of those in the $\lambda=\ydiagram{1}$ sector absent in the gravity sector---are related to each other:  once the deformation is turned on, they combine and lift.  Let us examine this in more detail.

Recall that the affine multiplets we identified in the gravity sector and the affine primaries in the full CFT listed above are BPS states of the free orbifold CFT\@.
It is possible for them to lift once a deformation is turned on. 
To study the possibility of such lifting, however, neither the BPS partition function \eqref{eq:SW_PF} nor its contributions from individual symmetry sectors~\eqref{eq:S_CFT_exp_refined} is an appropriate quantity, since neither is protected.
A more suitable quantity is the resolved elliptic genus (REG)~\cite{Hughes:2026qqn}: a protected supersymmetric index which provides a one-parameter generalisation of the more standard modified elliptic genus~\cite{Maldacena:1999bp}.
While the general expression of the REG, a generating function for which can be found in \cite[Eq.~(5.11)]{Hughes:2026qqn}, is not necessary for the present work, for the $N=2$ theory it simply gives two separately protected indices given by the following weighted sums of left-moving symmetry sector contributions:
\begin{subequations}\label{eq:REG_N=2_def} 
\begin{align}
    \cE_{2,0}(q,y) &= 
     2S_{\ydiagram{2}}(q,y,1,1) 
    +
    S_{\YTscriptsize\ydiagram{1}}(q,y,1,1) 
    \ ,\label{eq:REG_N=2_def_1} \\
    \cE_{2,\frac12}(q,y) &= -2S_{\ydiagram{1,1}}(q,y,1,1) \ ,\label{eq:REG_N=2_def_2}
\end{align}
\end{subequations}
where the $S_{\lam}$ are given in \eqref{eq:Schur_poly_defs}. In other words, the $\lam=\YTnormalsize\ydiagram{1,1}$ sector is protected by itself (there is no lifting involving these states), whereas only a particular combination of the $\lam=\ydiagram{1}$ and $\lam=\ydiagram{2}$ sectors yields a protected quantity.

Substituting the CFT partition function \eqref{eq:S_CFT_exp_refined} into the REGs \eqref{eq:REG_N=2_def}, we obtain the following decompositions in contracted large $\cN=4$ characters:
\begin{subequations} \label{eq:REG_exp_refined}
    \begin{align}
        \cE_{2,0}^{\rm CFT} &= 2\mathbf{\Phi}_{0} + \mathbf{\Phi}_{\frac12} - 26 \mathbf{\Phi}_{0,2} - 62 \mathbf{\Phi}_{0,3} + \cdots \ , \label{eq:REG_exp_refined_1}\\
        \cE_{2,\frac12}^{\rm CFT} &= 4 \mathbf{\Phi}_{\frac12} - 16 \mathbf{\Phi}_{0,2} -8 \mathbf{\Phi}_{0,3} + \cdots \ . \label{eq:REG_exp_refined_2}
    \end{align}
\end{subequations}
By comparison of the first REG sector~\eqref{eq:REG_exp_refined_1} with the original partition functions \eqref{eq:S_CFT_2_exp_refined} and \eqref{eq:S_CFT_1_exp_refined}, we see that there have been a number of cancellations between affine characters from the $\lam=\YTnormalsize\ydiagram{2}\,$ and $\lam=\ydiagram{1}\,$ symmetry sectors, indicative of multiplets combining and lifting: \textit{i.e.}, becoming non-BPS in the theory at a generic point in its moduli space. In fact, all of the long affine characters in the $\lambda=\ydiagram{2}$ sector have dropped out of the index, due to cancellations with characters in the $\lambda=\ydiagram{1}$ sector. This observation agrees with the direct lifting calculations of \cite{Guo:2019ady,Guo:2020gxm}, in which all lifted states in the $N=2$ theory were identified up to level 4.

Explicitly, the combinations of multiplets involved in the lifting are given (in terms of their affine primary states\footnote{Since the left-moving affine algebra (anti-)commutes with the deformed right-moving supercharge, responsible for generating lifted quartets of states, the lifting of affine multiplets is determined by the lifting of its affine primary.}) by
\begin{subequations} \label{eq:lifted_quartets}
\begin{align}
    (j,h)=(0,1) &: \qquad \big| a^{(\rm{II})}_{0,1} \big>^a_{\ydiagram{2}}\, \big|\tilde{\phi}^{(1)}\big>_{\ydiagram{2}} 
    \xlongrightarrow{\ \tilde{\cG}^{\dot{+}A}_{\!-\frac12}\ } 
    \ket{\mathfrak{a}_{0,1}}^{A,a}_{\ydiagram{1}} \big|\tilde{\phi}\big> _{\YTscriptsize\ydiagram{1}}
    \xlongrightarrow{\, \ep_{AB}\tilde{\cG}^{\dot{+}B}_{\!-\frac12}\, } 
    \big| a^{(\rm{II})}_{0,1} \big>^a_{\ydiagram{2}} \big|\tilde{\phi}^{(2)}\big>_{\ydiagram{2}}\ , \label{eq:lifted_quartet_1}\\[1ex]
    (j,h)=(0,2) &: \qquad 
    \big| a^{(\rm{III})}_{0,2} \big>_{\ydiagram{2}}\,
    \big|\tilde{\phi}^{(1)}\big>_{\ydiagram{2}}
    \xlongrightarrow{\ \tilde{\cG}^{\dot{+}A}_{\!-\frac12}\ } 
    \ket{\mathfrak{a}_{0,2}}^{A}_{\ydiagram{1}} \big|\tilde{\phi}\big>_{\YTscriptsize\ydiagram{1}} \xlongrightarrow{\, \ep_{AB}\tilde{\cG}^{\dot{+}B}_{\!-\frac12}\, }
    \big| a^{(\rm{III})}_{0,2} \big>_{\ydiagram{2}} \big|\tilde{\phi}^{(2)}\big>_{\YTscriptsize\ydiagram{2}} \ , 
    \label{eq:lifted_quartet_2}
\end{align}
\end{subequations}
where $\tilde{\cG}^{\dot{\pm}A}_{\!\mp\frac12}$ are the deformed right-moving supercharges which generate lifted quartets, as detailed in~\cite{Guo:2019pzk,Hughes:2026qqn}. Above we have also included the right-moving parts of states in order to show the lifted quartet structure, with $\ket*{\tilde{\phi}}_{\YTscriptsize\ydiagram{1}}$ representing the 4 states in \eqref{eq:right-moving_states_1} and $\ket*{\tilde{\phi}^{(1,2)}}_{\YTscriptsize\ydiagram{2}}$ representing half each of the 8 states in \eqref{eq:right-moving_states_2}.\footnote{The right-moving fermion total modes $\tilde{\psi}^{\dot{\pm}\Ad(\rm T)}_{\mp\frac12}$ generate $SO(4)$ Clifford quartets of right-chiral states and also anti-commute with the deformed supercharges $\tilde{\cG}^{\dot{\pm}A}_{\!\mp\frac12}$. The right-moving states in the $\lam=\YTnormalsize\ydiagram{1}$ symmetry sector, given in \eqref{eq:right-moving_states_1}, form one such Clifford quartet. In the $\lam=\ydiagram{2}$ symmetry sector, the states in \eqref{eq:right-moving_states_2} decompose into an $SU(2)_R$ doublet of Clifford quartets, represented in \eqref{eq:lifted_quartets} by the states $\ket*{\tilde{\phi}^{(1)}}_{\YTscriptsize\ydiagram{2}}$ and $\ket*{\tilde{\phi}^{(2)}}_{\YTscriptsize\ydiagram{2}}$. Each right-moving Clifford quartet of states contributes 1 to the index, hence the relative coefficients in the REG \eqref{eq:REG_N=2_def_1}.} The states in \eqref{eq:lifted_quartets} are equivalent (up to normalisations and a spectral flow~\cite{Schwimmer:1986mf} to the Ramond-Ramond sector) with those in Eqs.~(6.2) and (6.4) of \cite{Guo:2020gxm}. 

Thus, all BPS long multiplet states in the gravity sector in the $\lambda=\YTnormalsize\ydiagram{2}$ symmetry sector (up to level two) combine with the stringy states in the $\lambda=\ydiagram{1}$ sector and lift, once the theory is deformed. As a result, no long multiplet remain in the $\lambda=\ydiagram{2}$ sector, whereas some stringy long multiplets in the $\lambda=\ydiagram{1}$ sector remain BPS, as do all the long multiplets in the $\lambda=\ydiagram{1,1}$ sector.
It is curious that all of the lifted quartets, up to the level we work at, involve mixing between multiplets within the gravity sector and multiplets outside of it. This shows that the gravity sector of the symmetric orbifold's BPS spectrum is not a closed subsector. We will comment on interpretations of these results in Section~\ref{sec:discussion}.

\section{Discussion} \label{sec:discussion}

% \begin{itemize}
%     \item conjecture that pattern continues to all levels for $N=2$, supported in the symmetric and twisted sectors by information in \cite{Chang:2025wgo}.
%     \item bulk interpretation?
%     \item scaling of lifting?
%     \item mention differences with higher $N$? de Boer bound etc
% \end{itemize}

In this paper we have proposed an algorithmic approach to constructing the gravity sector ($H^{\mathrm{gravity}}_0$) of $\mathrm{Sym}^N(T^4)$; that is, the completion of the graviton sector ($H^{\mathrm{graviton}}_0$) into multiplets of the symmetry algebra of the D1-D5 CFT (the contracted large $\cN=4$ algebra). We have carried out this procedure for the $N=2$ theory, explicitly constructing all affine primaries within the gravity sector up to level 2. We also constructed the affine primaries in the complement (within the space of free-theory BPS states~\eqref{eq:Hilbert_space_N}) Hilbert space $\overline{H^{\mathrm{gravity}}_0}$, up to the same level. By the use of a supersymmetric index and the calculations of \cite{Guo:2020gxm}, a number of the multiplets in $H^{\mathrm{gravity}}_0$ and $\overline{H^{\mathrm{gravity}}_0}$ are found to combine and lift in the interacting (second-order deformed) theory.

We now briefly comment on these results.
Firstly, we are now inclined to interpret the remaining (unlifted) BPS states in the language of the fortuity classification~\cite{Chang:2024zqi}. Since monotone states have a smooth large $N$ limit, and thus should be visible in the supergravity regime, the unlifted multiplets in the gravity sector should fall into this category. The remaining multiplets, those not in the supergravity sector and not lifted, might then reasonably be expected to fall into the fortuitous category. We summarise this expected classification of states, as well as the lifted states, in Table~\ref{tab2}. While the above picture is reasonable, it is of course necessary to back these statements up with explicit computation via the procedure outlined in \cite{Chang:2025rqy}.

With regards to our two proposed twisted-sector ($\lambda=\YTnormalsize\ydiagram{1}\,$) fortuitous multiplets at level~2, with affine primaries $\ket*{\mathfrak{a}'_{0,2}}^{ABC,\Ad\Bd\Cd\Dd}_{\YTscriptsize\ydiagram{1}}$ and $\ket*{\mathfrak{a}''_{0,2}}^{A,a}_{\YTscriptsize\ydiagram{1}}$ in \eqref{eq:APrim_long_5-7}, these correspond to the highest-weight fortuitous states presented in Equations (2.7) and (2.8) of~\cite{Chang:2025wgo}. On the other hand, the $\lam=\YTnormalsize\ydiagram{1,1}$ sector state $\ket*{\mathfrak{a}_{0,2}}^{A,\Ad\Bd\Cd}_{\YTscriptsize\ydiagram{1,1}}$ in \eqref{eq:APrim_long_3} has not, to our knowledge, appeared before in the literature. 
Our tentative perturbative analysis suggests that this state is likewise fortuitous: we expect that it is lifted once suitably embedded in the $\mathrm{Sym  }^{3}(T^4)$ theory. It would be desirable to confirm this expectation by an explicit calculation.

One point of note on our constructed gravity-sector states is that, while they exist within all symmetry sectors, long gravity multiplets are only found in the symmetric sector ($\lam=\YTnormalsize\ydiagram{2}$).
In turn, all of these long multiplets then lift. While it was known that lifting is only possible between the $\lam=\ydiagram{2}$ and $\lam=\ydiagram{1}$ symmetry sectors~\cite{Guo:2020gxm,Hughes:2026qqn} in the $N=2$ theory, and that all symmetric-sector long multiplets in the CFT lift up to level 4~\cite{Guo:2020gxm}, it was not known that these lifted states all belong to the gravity sector and none of them are stringy. This suggests a broader pattern: for $N=2$, all long multiplets in the $\lam=\ydiagram{2}$ symmetry sector are within the gravity sector, all of which lift by combining with non-gravity states from the $\lam=\ydiagram{1}$ sector. It is natural to further conjecture that this continues to be the only type of lifting in the $N=2$ theory.

Based on these results, we then propose the following: after taking into account lifting, the gravity sector Hilbert space $H^{\mathrm{gravity}}_0$ and its complement $\overline{H^{\mathrm{gravity}}_0}$ respectively reduce to the monotone and fortuitous Hilbert spaces, $H^{\mathrm{mon}}$ and $H^{\mathrm{for}}$, of~\cite{Chang:2025rqy}. This is supported by~\cite{Chang:2025wgo}, where it was stated that all monotone states up to level 4 fall into short multiplets. Our results are also consistent with the R-charge concentration phenomenon~\cite{Chang:2025wgo} which, for $N=2$, states that fortuitous states are only present in the $\lam=\ydiagram{1}$ and $\lam=\ydiagram{1,1}$ sectors. It would be interesting to examine whether this approach yields an alternative definition of monotone and fortuitous states for $N>2$ and the associated supersymmetry indices \cite{Hughes:2025tdy}; something we hope to examine in the near future. Another reason to consider larger values of $N$, in particular $N\geq5$, is that the first affine long multiplets (at $h=1,j=0$) fall below the black hole bound; the region of agreement between the CFT and supergraviton indices~\cite{Maldacena:1999bp,deBoer:1998us}. It would be interesting to see how this agreement emerges from an analysis of the gravity sector and its lifting for these cases.

The bulk interpretation of the observed lifted multiplets (in \eqref{eq:lifted_quartets}), involving mixing between gravity-sector states and states within the wider CFT spectrum, is unclear to us. Typically, lifted non-graviton states are thought of as stringy states due to their fractional mode excitations~\cite{Gava:2002xb}. This observed lifting seems fundamentally different in nature to the lifting of multi-particle (multi-trace) operators from $\frac1{N}$ effects observed from supergravity-regime correlation functions (for example in AdS$_3$~\cite{Giusto:2020mup,Ceplak:2021wzz,Aprile:2021mvq} and AdS$_5$~\cite{Aprile:2017xsp}), which occurs due to interactions between supergravity excitations. This observed lifting does, however, make it clear that the gravity sector is not a closed subsector of the CFT BPS spectrum.

In~\cite{Hughes:2025tdy} it was observed that a subset of the proposed fortuitous at $h=1,j=0$ in $\mathrm{Sym}^2(K3)$ are also valid states in $\mathrm{Sym}^2(T^4)$. An interesting question to ask is: what is the fate of these states in the $T^4$ theory? The normalised left-moving part of the states in question, adapted to our current conventions, are
\begin{equation} \label{eq:K3_For}
    \frac1{4\sqrt{2}}\Big[\ep_{\dot{C}\dot{D}}\psi^{-\dot{C}}_0\psi^{-\dot{D}}_0\ket*{\dot{A}}_2 + 2\sqrt{2}\psi^{-\dot{A}}_{-{1\over 2}}\ket{-}_2\Big] =  \frac18\Big[\ep_{\dot{C}\dot{D}}\psi^{-\dot{C}}_0\psi^{-\dot{D}}_0 \psi^{+\Ad}_{-\frac12} + 4\psi^{-\dot{A}}_{-{1\over 2}}\Big]\ket{-}_2 \ ,
\end{equation}
which is orthogonal to the global descendant $J^-_0\ket*{\Ad}_2$\@. The state \eqref{eq:K3_For} is a singleton global primary in tower $\mathrm{I}_{\YTscriptsize\ydiagram{1}}$ described in section~\ref{sssec:N=(2)} and is responsible for the contribution $-\dot{\chi}_\frac{1}{2}\phi^{(\rm I)}_{0,1}$ in \eqref{eq:towerItilde}. Under the relation between the gravity sector of this paper and the monotone sector proposed above, the state \eqref{eq:K3_For} is therefore fortuitous in $K3$ and monotone in $T^4$.  The reason for the difference is simple: in $T^4$, the state \eqref{eq:K3_For} is a singleton state and is thus monotone because of the additional global generator $\psi^{\alpha\Ad}$ which do not exist in $K3$.

\begin{table}[tb]
\centering
\begin{math}
\renewcommand{\arraystretch}{1.3}
\begin{array}{c||c|c|c}
%\hline
\raisebox{0.5ex}{\strut}(j,h)\raisebox{-0.5ex}{\strut} &\text{Monotone}&\text{Fortuitous}&\text{Lifted} \\
\hline
\hline
\raisebox{1ex}{\strut}(j,j)\raisebox{-1ex}{\strut} & \ket*{a_{0}^{(\rm I)}}_{\YTscriptsize\ydiagram{2}}, \ket*{a_{\frac12}^{({\rm I})}}_{\YTscriptsize\ydiagram{1}}, \ket*{a_{\frac12}^{(\rm I)}}^{\Ad}_{\YTscriptsize\ydiagram{1,1}} & \text{---} & \text{---}\\
\hline
\raisebox{1ex}{\strut}(0,1)\raisebox{-1ex}{\strut} & \text{---} & \text{---} & \ket*{a^{(\rm{II})}_{0,1}}^a_{\ydiagram{2}}, \ket*{\mathfrak{a}_{0,1}}^{A,a}_{\ydiagram{1}}\\
\hline
\raisebox{1ex}{\strut}(0,2)\raisebox{-1ex}{\strut} & \text{---} & \ket*{\mathfrak{a}_{0,2}}^{A,\Ad\Bd\Cd}_{\YTscriptsize\ydiagram{1,1}}, \ket*{\mathfrak{a}'_{0,2}}^{ABC,\Ad\Bd\Cd\Dd}_{\YTscriptsize\ydiagram{1}}, \ket*{\mathfrak{a}''_{0,2}}^{A,a}_{\YTscriptsize\ydiagram{1}} & \ket*{a^{(\rm{III})}_{0,2}}_{\ydiagram{2}}, \ket*{\mathfrak{a}_{0,2}}^{A}_{\ydiagram{1}}\\ 
\end{array}
\end{math}\YTnormalsize
\caption{\label{tab2} \sl The constructed affine primary states of the $N=2$ free theory BPS spectrum, sorted according to the proposed connection with the fortuity classification. It is expected that at higher levels, it continues to hold that all long affine multiplets within the gravity sector are in the $\lam=\ydiagram{2}$ symmetry sector and are lifted.}
\end{table}

%%%%%%%%%%%%%%%%%%%%%%%%%%%%%%%%%%%%%%%%%%%%%%%%%%%%%%%%%%%%%%%%%%%%
\section*{Acknowledgements}

We would like to thank Jan de Boer, Stefano Giusto, Rodolfo Russo and David Turton for fruitful discussions.
We thank ChatGPT (OpenAI) for assistance with Mathematica coding to implement $\cN=4$ algebra computations.
This work was supported in part by MEXT
KAKENHI Grant Numbers 21H05184 and 24K00626.

%%%%%%%%%%%%%%%%%%%%%%%%%%%%%%%%%%%%%%%%%%%%%%%%%%%%%%%%%%%%%%%%%%%%
\appendix

\section{Explicit primary states}
\label{app:expl_states}

In this appendix, we write down the explicit forms of global primaries that appeared in section~\ref{sec:results}. For some cases we also present details of the computations not presented in the main text.
Note that we do not keep track of the normalization of the states.

\YTnormalsize
\subsection{Untwisted symmetric sector ($\lambda=\ydiagram{2}\mkern1.5mu$)}
\subsubsection*{Graviton global primaries (except for $(j,h)=(0,2)$):}
%\begin{align}
%     &\big| g^{(\frac{1}{2})}_{\frac{1}{2}} \big>^{\dot{A}} = \ket{-}^{[1]}_1 \ket*{\dot{A}}^{[2]}_1 + \ket*{\dot{A}}^{[1]}_1 \ket{-}^{[2]}_1,\\
%     &\big| g^{(1)}_1 \big> = \epsilon_{\dot{A} \dot{B}} (\ket*{\dot{A}}^{[1]}_1 \ket*{\dot{B}}^{[2]}_1 - \ket*{\dot{B}}^{[1]}_1 \ket*{\dot{A}}^{[2]}_1),\\
%     &\big| g^{(1')}_1 \big> = \ket{-}^{[1]}_1 \ket{+}^{[2]}_1 + \ket{+}^{[1]}_1 \ket{-}^{[2]}_1,\\
%     &\big| g^{(1)}_{0,1} \big>^a = \sigma^a_{\dot{A} \dot{B}} (J^-_0 \ket*{\dot{A}}^{[1]}_1 \ket*{\dot{B}}^{[2]}_1 - \ket*{\dot{B}}^{[1]}_1 J^-_0 \ket*{\dot{A}}^{[2]}_1 ),\\
%     &\big| g^{(\frac{3}{2})}_{\frac{3}{2}} \big>^{\dot{A}} = \ket{+}^{[1]}_1 \ket*{\dot{A}}^{[2]}_1 + \ket*{\dot{A}}^{[1]}_1 \ket{+}^{[2]}_1,\\
%     &\big| g^{(\frac{3}{2})}_{\frac{1}{2},\frac{3}{2}} \big>^{\dot{A}} = \left( J^{-}_0 \ket{+}^{[1]}_1 \ket*{\dot{A}}^{[2]}_1 -2 \ket{+}^{[1]}_1 J^{-}_0 \ket*{\dot{A}}^{[2]}_1 \right) \notag\\
%     &\qquad\qquad\qquad\qquad+ \left( \ket*{\dot{A}}^{[1]}_1 J^{-}_0 \ket{+}^{[2]}_1 -2 J^{-}_0 \ket*{\dot{A}}^{[1]}_1 \ket{+}^{[2]}_1 \right),\\
%     &\big| g^{(2)}_2 \big> = \ket{+}^{[1]}_1 \ket{+}^{[2]}_1.
%\end{align}]

\begin{align}
     \big| g^{(\frac{1}{2})}_{\frac{1}{2}} \big>^{\dot{A}} &= \ket{-}^{[1]}_1 \ket*{\dot{A}}^{[2]}_1 + \ket*{\dot{A}}^{[1]}_1 \ket{-}^{[2]}_1,\\
     \big| g^{(1)}_1 \big> &= \epsilon_{\dot{A} \dot{B}} \ket*{\dot{A}}^{[1]}_1 \ket*{\dot{B}}^{[2]}_1,\\
     \big| g^{(1')}_1 \big> &= \ket{-}^{[1]}_1 \ket{+}^{[2]}_1 + \ket{+}^{[1]}_1 \ket{-}^{[2]}_1,\\
     \big| g^{(1)}_{0,1} \big>^a &=  \left(\sigma^a \right)_{\dot{A} \dot{B}} J^{-(\mathcal{A})}_0 \ket*{\dot{A}}^{[1]}_1 \ket*{\dot{B}}^{[2]}_1 ,\\
     \big| g^{(\frac{3}{2})}_{\frac{3}{2}} \big>^{\dot{A}} &= \ket{+}^{[1]}_1 \ket*{\dot{A}}^{[2]}_1 + \ket*{\dot{A}}^{[1]}_1 \ket{+}^{[2]}_1,\\
     \big| g^{(\frac{3}{2})}_{\frac{1}{2},\frac{3}{2}} \big>^{\dot{A}} &= \left( J^{-[1]}_0 \ket{+}^{[1]}_1 \ket*{\dot{A}}^{[2]}_1 -2 \ket{+}^{[1]}_1 J^{-[2]}_0 \ket*{\dot{A}}^{[2]}_1 \right) \notag\\[-1ex]
     &\qquad\qquad\qquad+ \left( \ket*{\dot{A}}^{[1]}_1 J^{-[2]}_0 \ket{+}^{[2]}_1 -2 J^{-[1]}_0 \ket*{\dot{A}}^{[1]}_1 \ket{+}^{[2]}_1 \right),\\
     \big| g^{(2)}_2 \big> &= \ket{+}^{[1]}_1 \ket{+}^{[2]}_1.
\end{align}

\subsubsection*{Gravity towers}
Affine primaries:
\begin{align}
    \big| a^{(\rm{I})}_{0} \big> &= \ket{-}^{[1]}_1 \ket{-}^{[2]}_1,\\
    \big| a^{(\rm{II})}_{0,1} \big>^a &=-\frac{1}{2} \left(\sigma^a \right)_{\dot{B}\dot{C}} \psi^{+\dot{A}(\rm{T})}_{-\frac{1}{2}} \psi^{-\dot{B}(\mathcal{A})}_{-\frac{1}{2}} \psi^{+\dot{C}(\mathcal{A)}}_{-\frac{1}{2}} \ket{-}^{[1]}_1 \ket{-}^{[2]}_1.
\end{align}

\noindent
Singleton global primaries (except for $(j,h)=(0,2)$):
\begin{align}
    \big| s^{(\rm{I})}_{\frac{1}{2}} \big>^{\dot{A}} &= \psi^{+\dot{A}(\rm{T})}_{-\frac{1}{2}} \ket{-}^{[1]}_1 \ket{-}^{[2]}_1,\\
    \big| s^{(\rm{I})}_1 \big> &=\Big( J^{+(\rm{T})}_{-1} - \frac{1}{2} \epsilon_{\dot{A} \dot{B}} \psi^{+\dot{A}(\rm{T})}_{-\frac{1}{2}} \psi^{+\dot{B}(\rm{T})}_{-\frac{1}{2}} \Big) \ket{-}^{[1]}_1 \ket{-}^{[2]}_1,\\
    \big| s'^{(\rm{I})}_1 \big> &= J^{+(\rm{T})}_{-1} \ket{-}^{[1]}_1 \ket{-}^{[2]}_1,\\
    \big| s^{(\rm{I})}_{0,1} \big>^a &= \left(\sigma^a \right)_{\dot{A}\dot{B}} \psi^{-\dot{A}(\rm{T})}_{-\frac{1}{2}} \psi^{+\dot{B}(\rm{T})}_{-\frac{1}{2}} \ket{-}^{[1]}_1 \ket{-}^{[2]}_1,\\
    \big| s^{(\rm{I})}_{\frac{3}{2}} \big>^{\dot{A}} &= \psi^{+\dot{A}(\rm{T})}_{-\frac{1}{2}} J^{+(\rm{T})}_{-1} \ket{-}^{[1]}_1 \ket{-}^{[2]}_1,\\
    \big| s^{(\rm{I})}_{\frac{1}{2},\frac{3}{2}} \big>^{\dot{A}} &=\frac{1}{2} \epsilon_{\dot{B}\dot{C}} \psi_{-\frac{1}{2}}^{-\dot{A}(\rm{T})} \psi_{-\frac{1}{2}}^{+\dot{B}(\rm{T})} \psi_{-\frac{1}{2}}^{+\dot{C}(\rm{T})} \ket{-}^{[1]}_1 \ket{-}^{[2]}_1,\\
    \big| s'^{(\rm{I})}_{\frac{1}{2},\frac{3}{2}} \big>^{\dot{A}} &=\left( \psi_{-\frac{3}{2}}^{+\dot{A}(\rm{T})} -\frac{2}{3} J^{+(\rm{T})}_{-1} \psi_{-\frac{1}{2}}^{-\dot{A}(\rm{T})} -\frac{2}{3} J^{3(\rm{T})}_{-1} \psi_{-\frac{1}{2}}^{+\dot{A}(\rm{T})} \right.\notag \\[-1ex]
    &\left.\qquad\qquad\qquad\qquad\quad+ \frac{1}{4} \epsilon_{\dot{A}\dot{B}} \psi_{-\frac{1}{2}}^{-\dot{A}(\rm{T})} \psi_{-\frac{1}{2}}^{+\dot{B}(\rm{T})} \psi_{-\frac{1}{2}}^{+\dot{C}(\rm{T})} \right) \ket{-}^{[1]}_1 \ket{-}^{[2]}_1,\\
    \big| s^{(\rm{II})}_{\frac{1}{2},\frac{3}{2}} \big>^{\dot{A},a} &=\psi^{+\dot{A}(\rm{T})}_{-\frac{1}{2}} \big| a^{(\rm{II})}_{0,1} \big>^a=-\frac{1}{2} \left(\sigma^a \right)_{\dot{B}\dot{C}} \psi^{+\dot{A}(\rm{T})}_{-\frac{1}{2}} \psi^{-\dot{B}(\mathcal{A})}_{-\frac{1}{2}} \psi^{+\dot{C}(\mathcal{A)}}_{-\frac{1}{2}} \ket{-}^{[1]}_1 \ket{-}^{[2]}_1,\\
    \big| s^{(\rm{I})}_{2} \big> &= \frac{1}{2} \big( J^{+(\rm{T})}_{-1} \big)^2 \ket{-}^{[1]}_1 \ket{-}^{[2]}_1.
\end{align}

\subsubsection*{The case of $(j,h)=(0,2)$}

For these charges we present details not presented in the main text.

The graviton global primaries corresponding to $\phi^{(1)}_{0,2}$ in (\ref{eq:1}) and $\phi^{(2)}_{0,2}$ in (\ref{eq:2}) are
%\begin{align}
%    &\big| g^{(1)}_{0,2} \big> \notag\\
%    &= \left( L_{-1} J_0^{-}  \ket*{\dot{A}}^{[1]}_1 \ket*{\dot{B}}^{[2]}_1 + \ket*{\dot{A}}^{[1]}_1 L_{-1} J_0^{-} \ket*{\dot{B}}^{[2]}_1 + L_{-1} \ket*{\dot{A}}^{[1]}_1 J_0^{-} \ket*{\dot{B}}^{[2]}_1 + J_0^{-} \ket*{\dot{A}}^{[1]}_1 L_{-1} \ket*{\dot{B}}^{[2]}_1 \right. \notag\\
%    &\left. \qquad+ 2 G_{-\frac{1}{2}}^{-1} G_{-\frac{1}{2}}^{-2} \ket*{\dot{A}}^{[1]}_1 \ket*{\dot{B}}^{[2]}_1 + 2 \ket*{\dot{A}}^{[1]}_1 G_{-\frac{1}{2}}^{-1}  G_{-\frac{1}{2}}^{-2} \ket*{\dot{B}}^{[2]}_1 + \epsilon_{AB} G_{-\frac{1}{2}}^{-A} \ket*{\dot{A}}^{[1]}_1 G_{-\frac{1}{2}}^{-B} \ket*{\dot{B}}^{[2]}_1 \right) \epsilon_{\dot{A} \dot{B}},\\
%    &\big| g^{(2)}_{0,2} \big> = J_0^- \ket{+}^{[1]}_1 J_0^- \ket{+}^{[2]}_1 - (J_0^-)^2 \ket{+}^{[1]}_1 \ket{+}^{[2]}_1 - \ket{+}^{[1]}_1 (J_0^-)^2 \ket{+}^{[2]}_1.
%\end{align}
\begin{align}
    \big| g^{(1)}_{0,2} \big>
    &=  \Bigl[ L_{-1}^{(\rm{T})} J_0^{-(\rm{T})}
    \notag\\[-1ex]
    &\qquad + \epsilon_{AB} \left(G_{-\frac{1}{2}}^{-A[1]}G_{-\frac{1}{2}}^{-B[1]}+G_{-\frac{1}{2}}^{-A[2]}G_{-\frac{1}{2}}^{-B[2]}-G_{-\frac{1}{2}}^{-A[1]}G_{-\frac{1}{2}}^{-B[2]}\right) \Bigr] \epsilon_{\dot{A} \dot{B}} \ket*{\dot{A}}^{[1]}_1 \ket*{\dot{B}}^{[2]}_1 ,\\
    \big| g^{(2)}_{0,2} \big> &= \left(J_0^{-[1]} J_0^{-[2]} - (J_0^{-[1]})^2 - (J_0^{-[2]})^2 \right) \ket{+}^{[1]}_1 \ket{+}^{[2]}_1.
\end{align}
The singleton global primaries corresponding to $4 \phi^{(\rm{I})}_{0,2}$ in (\ref{eq:tower1}), taken mutually orthogonal, are
%\begin{align}
%    &\big| s^{(\rm{I})}_{0,2} \big> = \ket{4} =\psi^{+ \dot{1}}_{-\frac{1}{2}} \psi^{+ \dot{2}}_{-\frac{1}{2}} \psi^{- \dot{1}}_{-\frac{1}{2}} \psi^{- \dot{2}}_{-\frac{1}{2}} (\ket{-}_1 \ket{-}_1),\\
%    &\big| s'^{(\rm{I})}_{0,2} \big>=\ket{1}-\frac{3}{4} \ket{4} ,\\
%    &\big| s''^{(\rm{I})}_{0,2} \big>=\ket{2}-\frac{8}{11}\big| s'^{(\rm{I})}_{0,2} \big> ,\\
%    &\big| s'''^{(\rm{I})}_{0,2} \big>=\ket{3}-3\ket{4}-\frac{8}{11}\big| s'^{(\rm{I})}_{0,2} \big>+\frac{2}{9} \big| s''^{(\rm{I})}_{0,2} \big>,
%\end{align}
\begin{align}
    \big| s^{(\rm{I})}_{0,2} \big> &= \ket{4} =\psi^{+ \dot{1}(\rm{T})}_{-\frac{1}{2}} \psi^{+ \dot{2}(\rm{T})}_{-\frac{1}{2}} \psi^{- \dot{1}(\rm{T})}_{-\frac{1}{2}} \psi^{- \dot{2}(\rm{T})}_{-\frac{1}{2}} (\ket{-}_1 \ket{-}_1),\\
    \big| s'^{(\rm{I})}_{0,2} \big>&=\ket{1}-\frac{3}{4} \ket{4} ,\\
    \big| s''^{(\rm{I})}_{0,2} \big>&=\ket{2}-\frac{8}{11}\big| s'^{(\rm{I})}_{0,2} \big> ,\\
    \big| s'''^{(\rm{I})}_{0,2} \big>&=\ket{3}-3\ket{4}-\frac{8}{11}\big| s'^{(\rm{I})}_{0,2} \big>+\frac{2}{9} \big| s''^{(\rm{I})}_{0,2} \big>,
\end{align}
where
%\begin{align}
%    &\ket{1}=\left[ 2J_{-2}^3 -2(J_{-1}^3)^2 -2 J_{-1}^+ J_{-1}^- +L_{-2} \right] (\ket{-}^{[1]}_1 \ket{-}^{[2]}_1),\\
%    &\ket{2}=\left[ \epsilon_{\dot{A} \dot{B}} \epsilon_{A B} \alpha^{A \dot{A}}_{-1} \alpha^{B \dot{B}}_{-1} - 2\epsilon_{\dot{A} \dot{B}} (\psi_{-\frac{1}{2}}^{+\dot{A}} \psi_{-\frac{3}{2}}^{-\dot{B}} - \psi_{-\frac{1}{2}}^{-\dot{A}} \psi_{-\frac{3}{2}}^{+\dot{B}})\right](\ket{-}^{[1]}_1 \ket{-}^{[2]}_1),\\
%    &\ket{3} =\left[ \epsilon_{\dot{A} \dot{B}} (J^+_{-1} \psi^{-\dot{A}}_{-\frac{1}{2}} \psi^{-\dot{B}}_{-\frac{1}{2}} - J^-_{-1} \psi^{+\dot{A}}_{-\frac{1}{2}} \psi^{+\dot{B}}_{-\frac{1}{2}} + 2J^3_{-1} \psi^{+\dot{A}}_{-\frac{1}{2}} \psi^{-\dot{B}}_{-\frac{1}{2}}) \right. \notag\\
%    &\left. \qquad\qquad\qquad\qquad\qquad\qquad+ 3 \epsilon_{\dot{A} \dot{B}} (\psi_{-\frac{1}{2}}^{+\dot{A}} \psi_{-\frac{3}{2}}^{-\dot{B}} - \psi_{-\frac{1}{2}}^{-\dot{A}} \psi_{-\frac{3}{2}}^{+\dot{B}}) \right](\ket{-}^{[1]}_1 \ket{-}^{[2]}_1),\\
%    &\ket{4}=\psi^{+ \dot{1}}_{-\frac{1}{2}} \psi^{+ \dot{2}}_{-\frac{1}{2}} \psi^{- \dot{1}}_{-\frac{1}{2}} \psi^{- \dot{2}}_{-\frac{1}{2}} (\ket{-}^{[1]}_1 \ket{-}^{[2]}_1).
%\end{align}
\begin{align}
    \ket{1}&=\left[ 2J_{-2}^{3(\rm{T})} -2(J_{-1}^{3(\rm{T})})^2 -2 J_{-1}^{+(\rm{T})} J_{-1}^{-(\rm{T})} +L_{-2}^{(\rm{T})} \right]  \ket{-}^{[1]}_1 \ket{-}^{[2]}_1 ,\\
    \ket{2}&=\left[ \epsilon_{\dot{A} \dot{B}} \epsilon_{A B} \alpha^{A \dot{A}(\rm{T})}_{-1} \alpha^{B \dot{B}(\rm{T})}_{-1} - 2\epsilon_{\dot{A} \dot{B}} (\psi_{-\frac{1}{2}}^{+\dot{A}(\rm{T})} \psi_{-\frac{3}{2}}^{-\dot{B}(\rm{T})} - \psi_{-\frac{1}{2}}^{-\dot{A}(\rm{T})} \psi_{-\frac{3}{2}}^{+\dot{B}(\rm{T})})\right] \ket{-}^{[1]}_1 \ket{-}^{[2]}_1 ,\\
    \ket{3} &=\left[ \epsilon_{\dot{A} \dot{B}} (J^{+(\rm{T})}_{-1} \psi^{-\dot{A}(\rm{T})}_{-\frac{1}{2}} \psi^{-\dot{B}(\rm{T})}_{-\frac{1}{2}} - J^{-(\rm{T})}_{-1} \psi^{+\dot{A}(\rm{T})}_{-\frac{1}{2}} \psi^{+\dot{B}(\rm{T})}_{-\frac{1}{2}} + 2J^{3(\rm{T})}_{-1} \psi^{+\dot{A}(\rm{T})}_{-\frac{1}{2}} \psi^{-\dot{B}(\rm{T})}_{-\frac{1}{2}}) \right. \notag\\
    &\left. \qquad\qquad\qquad\qquad\qquad\;\;\;+ 3 \epsilon_{\dot{A} \dot{B}} (\psi_{-\frac{1}{2}}^{+\dot{A}(\rm{T})} \psi_{-\frac{3}{2}}^{-\dot{B}(\rm{T})} - \psi_{-\frac{1}{2}}^{-\dot{A}(\rm{T})} \psi_{-\frac{3}{2}}^{+\dot{B}(\rm{T})}) \right] \ket{-}^{[1]}_1 \ket{-}^{[2]}_1 ,\\
    \ket{4}&=\psi^{+ \dot{1}(\rm{T})}_{-\frac{1}{2}} \psi^{+ \dot{2}(\rm{T})}_{-\frac{1}{2}} \psi^{- \dot{1}(\rm{T})}_{-\frac{1}{2}} \psi^{- \dot{2}(\rm{T})}_{-\frac{1}{2}} \ket{-}^{[1]}_1 \ket{-}^{[2]}_1.
\end{align}
The singleton global primary corresponding to the $SU(2)_2$ singlet in the term $\dot{\chi}_1 \cdot \dot{\chi}_1 \phi^{(\rm{II})}_{0,2}$ in \eqref{eq:tower2} is
%\begin{align}
%    \big| s^{(\rm{II})}_{0,2} \big> =\delta_{ab} \sigma^a_{\dot{A} \dot{B}} \psi^{-\dot{A}}_{-\frac{1}{2}} \psi^{+\dot{B}}_{-\frac{1}{2}} \big| a^{(\rm{II})}_{0,1} \big>^b.
%\end{align}
\begin{align}
    \big| s^{(\rm{II})}_{0,2} \big> =\delta_{ab} \left(\sigma^a \right)_{\dot{A} \dot{B}} \psi^{-\dot{A}(\rm{T})}_{-\frac{1}{2}} \psi^{+\dot{B}(\rm{T})}_{-\frac{1}{2}} \big| a^{(\rm{II})}_{0,1} \big>^b.
\end{align}
Thus, orthogonalizing the graviton global primaries against towers $\rm{I}$ and $\rm{II}$, we find 
\begin{subequations}
\label{ncpk15Apr26}
\begin{align}
    \big| g^{(1)}_{0,2} \big>_{\perp} &=\big| g^{(1)}_{0,2} \big> - \frac{11}{36} \ket{2} -\frac{1}{5} \ket{3},\\
    \big| g^{(2)}_{0,2} \big>_{\perp} &=\big| g^{(2)}_{0,2} \big> + \frac{3}{8} \big| s^{(\rm{I})}_{0,2} \big> +\frac{5}{22} \big| s'^{(\rm{I})}_{0,2} \big> -\frac{5}{72} \big| s''^{(\rm{I})}_{0,2} \big> - \frac{1}{40} \big| s'''^{(\rm{I})}_{0,2} \big> + \frac{3}{8} \big| s^{(\rm{II})}_{0,2} \big>.
\end{align}
\end{subequations}
In fact, these do not lead to independent states since
\begin{align}
      \frac{1}{2} \big| g^{(1)}_{0,2} \big>_{\perp}+\big| g^{(2)}_{0,2} \big>_{\perp} = 0.
\end{align}
We can therefore take $\big| g^{(1)}_{0,2} \big>_{\perp}$ as the independent state, which can be shown to be affine primary.
      
% Since they are not orthogonal to each other, orthogonalizing $\ket*{g'^{(2)}_{0,2}}_\perp$ to $\ket*{g^{(1)}_{0,2}}_\perp$, we find
% \begin{align}
%     \big| g^{(2)}_{0,2} \big>_{\perp} = \big| g'^{(2)}_{0,2} \big>_{\perp}+\frac{1}{2} \big| g^{(1)}_{0,2} \big>_{\perp} = 0.
% \end{align}

\subsection{Twisted sector ($\lambda=\ydiagram{1}\mkern1.5mu$)}
\subsubsection*{Graviton global primaries}
\begin{align}
    \ket*{g^{(0)}_\frac{1}{2}} =\ket{-}_2,\qquad
    \ket*{g^{(1)}_1}^{\dot{A}} =\ket*{\dot{A}}_2,\qquad
    \ket*{g^{(1)}_{\frac{2}{3}}} =\ket{+}_2.
\end{align}

\subsubsection*{Gravity towers}
Affine primary:
\begin{align}
    \big| a^{(\rm{I})}_{0} \big> &= \ket{-}_2.
\end{align}
Singleton global primaries:
\begin{align}
    \ket*{s^{(1)}_1}^{\dot{A}} &=\psi^{+\dot{A}(\rm{T})}_{-\frac{1}{2}}\ket{-}_2,\qquad
    \ket*{s^{(1)}_{\frac{3}{2}}} =J^{+(\rm{T})}_{-1}\ket{-}_2.
\end{align}

\subsection{Untwisted anti-symmetric sector ($\lambda=\ydiagram{1,1}\mkern1.5mu$)}
\subsubsection*{Graviton global primaries (except for $(j,h)$=(0,2)):}
\begin{align}
     \big| g^{(\frac{1}{2})}_{\frac{1}{2}} \big>^{\dot{A}} &= \ket{-}^{[1]}_1 \ket*{\dot{A}}^{[2]}_1 - \ket*{\dot{A}}^{[1]}_1 \ket{-}^{[2]}_1,\\
     \big| g^{(1)}_1 \big>^a &= \left(\sigma^a\right)_{\dot{A} \dot{B}} (\ket*{\dot{A}}^{[1]}_1 \ket*{\dot{B}}^{[2]}_1 + \ket*{\dot{B}}^{[1]}_1 \ket*{\dot{A}}^{[2]}_1),\\
     \big| g^{(1')}_1 \big> &= \ket{-}^{[1]}_1 \ket{+}^{[2]}_1 - \ket{+}^{[1]}_1 \ket{-}^{[2]}_1,\\
     \big| g^{(1)}_{0,1} \big> &= \epsilon_{\dot{A} \dot{B}} (J^{-[1]}_0 \ket*{\dot{A}}^{[1]}_1 \ket*{\dot{B}}^{[2]}_1 + \ket*{\dot{B}}^{[1]}_1 J^{-[2]}_0 \ket*{\dot{A}}^{[2]}_1 ),\\
     \big| g^{(\frac{3}{2})}_{\frac{3}{2}} \big>^{\dot{A}} &= \ket{+}^{[1]}_1 \ket*{\dot{A}}^{[2]}_1 - \ket*{\dot{A}}^{[1]}_1 \ket{+}^{[2]}_1,\\
     \big| g^{(\frac{3}{2})}_{\frac{1}{2},\frac{3}{2}} \big>^{\dot{A}} &= \left( J^{-[1]}_0 \ket{+}^{[1]}_1 \ket*{\dot{A}}^{[2]}_1 -2 \ket{+}^{[1]}_1 J^{-[2]}_0 \ket*{\dot{A}}^{[2]}_1 \right) \notag\\[-1ex]
     &\qquad\qquad\qquad\qquad- \left( \ket*{\dot{A}}^{[1]}_1 J^{-[2]}_0 \ket{+}^{[2]}_1 -2 J^{-[1]}_0 \ket*{\dot{A}}^{[1]}_1 \ket{+}^{[2]}_1 \right),\\
     \big| g^{(2)}_{1,2} \big> &= J^{-[1]}_0\ket{+}^{[1]}_1 \ket{+}^{[2]}_1-\ket{+}^{[1]}_1 J^{-[2]}_0 \ket{+}^{[2]}_1.
\end{align}

\subsubsection*{Gravity towers}

Affine primaries:
\begin{align}
    \big| a^{(\rm{I})}_{\frac{1}{2}} \big>_{\YTscriptsize\ydiagram{1,1}}^{\dot{A}}=(\ket{-}_1 \ket*{\dot{A}}_1)^{\YTscriptsize \ydiagram{1,1}}
    \equiv \ket{-}_1^{[1]} \ket*{\dot{A}}_1^{[2]}- \ket*{\dot{A}}_1^{[1]}\ket{-}_1^{[2]}.
\end{align}
Singleton global primaries  (except for $(j,h)$=(0,2)):
\begin{align}
    \big| s^{(\rm{I})}_1 \big>^a &= \left(\sigma^a\right)_{\dot{A}\dot{B}} \psi^{+\dot{A}(\rm{T})}_{-\frac{1}{2}} (\ket{-}_1 \ket*{\dot{B}}_1)^{\ytableausetup{boxsize=1.2mm} \ydiagram[ ]{1,1}},\\
    \big| s'^{(\rm{I})}_1 \big> &=\frac{1}{2}\epsilon_{\dot{A}\dot{B}} \psi^{+\dot{A}(\rm{T})}_{-\frac{1}{2}} (\ket{-}_1 \ket*{\dot{B}}_1)^{\ytableausetup{boxsize=1.2mm} \ydiagram[ ]{1,1}},\\
    \big| s^{(\rm{I})}_{0,1} \big> &= \epsilon_{\dot{A}\dot{B}} \left(\psi^{-\dot{A}(\rm{T})}_{-\frac{1}{2}} - \psi^{+\dot{A}(\rm{T})}_{-\frac{1}{2}} J^{-(\rm{T})}_0 \right)(\ket{-}_1 \ket*{\dot{B}}_1)^{\ytableausetup{boxsize=1.2mm} \ydiagram[ ]{1,1}},\\
    \big| s^{(\rm{I})}_{\frac{3}{2}} \big>^{\dot{A}} &= \frac{1}{4} \epsilon_{\dot{A}\dot{B}} \psi^{+\dot{A}(\rm{T})}_{-\frac{1}{2}} \psi^{+\dot{B}(\rm{T})}_{-\frac{1}{2}} (\ket{-}_1 \ket*{\dot{B}}_1)^{\ytableausetup{boxsize=1.2mm} \ydiagram[ ]{1,1}},\\
    \big| s^{(\rm{I})}_{\frac{1}{2},\frac{3}{2}} \big>^{\dot{A}} &=\left(\epsilon_{\dot{B}\dot{C}} \psi_{-\frac{1}{2}}^{+\dot{B}(\rm{T})} \psi_{-\frac{1}{2}}^{-\dot{C(\rm{T})}} +4J^{3(\rm{T})}_{-1} -2L_{-1}^{(\rm{T})}\right) (\ket{-}_1 \ket*{\dot{A}}_1)^{\ytableausetup{boxsize=1.2mm} \ydiagram[ ]{1,1}},\\
    \big| s'^{(\rm{I})}_{\frac{1}{2},\frac{3}{2}} \big>^{\dot{A}} &=\left(-2J^{3(\rm{T})}_{-1} +\frac{3}{4}\epsilon_{AB}G^{+A(\rm{T})}_{-\frac{1}{2}}G^{-B(\rm{T})}_{-\frac{1}{2}} +J^{+(\rm{T})}_{-1}J^{-(\rm{T})}_{0} \right. \notag\\[-1ex]
    &\left.\qquad\qquad\qquad\qquad-\frac{3}{4}\epsilon_{\dot{B}\dot{C}} \psi_{-\frac{1}{2}}^{+\dot{B}(\rm{T})} \psi_{-\frac{1}{2}}^{-\dot{C(\rm{T})}} +\frac{3}{2}L_{-1}^{(\rm{T})}\right) (\ket{-}_1 \ket*{\dot{A}}_1)^{\ytableausetup{boxsize=1.2mm} \ydiagram[ ]{1,1}},\\
    \big| s''^{(\rm{I})}_{\frac{1}{2},\frac{3}{2}} \big>^{\dot{A}}&=\sum_{a=1}^3 \left(\sigma^a\right)^{\dot{A}}_{\ \,\dot{B}} \left(\sigma^a\right)_{\dot{C}\dot{D}} \psi^{+\dot{C}(\rm{T})}_{-\frac{1}{2}} \psi^{-\dot{D}(\rm{T})}_{-\frac{1}{2}} (\ket{-}_1 \ket*{\dot{B}}_1)^{\ytableausetup{boxsize=1.2mm}\ydiagram[ ]{1,1}},\\
    \big| s^{(\rm{I})}_{1,2} \big> &=\epsilon_{\dot{A}\dot{B}} \left(J^{3(\rm{T})}_{-1} \psi^{+\dot{A}(\rm{T})}_{-\frac{1}{2}} +\frac{1}{4}\epsilon_{AB}G^{+A(\rm{T})}_{-\frac{1}{2}} G^{-B(\rm{T})}_{-\frac{1}{2}} \psi^{+\dot{A}}_{-\frac{1}{2}}\right) (\ket{-}_1 \ket*{\dot{B}}_1)^{\ytableausetup{boxsize=1.2mm}\ydiagram[ ]{1,1}},\\
    \big| s'^{(\rm{I})}_{1,2} \big> &=\epsilon_{\dot{A}\dot{B}} \left(J^{+(\rm{T})}_{-1} J^{-(\rm{T})}_0 \psi^{+\dot{A}(\rm{T})}_{-\frac{1}{2}} -2J^{3(\rm{T})}_{-1} \psi^{+\dot{A}(\rm{T})}_{-\frac{1}{2}}\right) (\ket{-}_1 \ket*{\dot{B}}_1)^{\ytableausetup{boxsize=1.2mm}\ydiagram[ ]{1,1}}.
\end{align}

\subsubsection*{The case of $(j,h)=(0,2)$}
The graviton global primaries corresponding to $\dot{\chi}_{1}\phi_{0,2}^{(1)}$ in \eqref{eq:1anti} are 
\begin{align}
    &\big| g^{(2)}_{0,2} \big>^a = (\sigma^a)_{\dot{A}\dot{B}}\left[ 2\left(L_{-1}^{[1]} \ket*{\dot{A}}^{[1]}_1 J_0^{-[2]} \ket*{\dot{B}}^{[2]}_1 + J_0^{-[1]} \ket*{\dot{A}}^{[1]}_1 L_{-1}^{[2]} \ket*{\dot{B}}^{[2]}_1 \right)\right.\notag\\
     &\qquad\qquad\qquad\qquad-2\left(L_{-1}^{[1]} J_0^{-[1]} \ket*{\dot{A}}^{[1]}_1 \ket*{\dot{B}}^{[2]}_1 + \ket*{\dot{A}}^{[1]}_1 L_{-1}^{[2]} J_0^{-[2]} \ket*{\dot{B}}^{[2]}_1\right)\notag\\
     &\left.\qquad\qquad\qquad\qquad+ \epsilon_{AB} \left(G^{-A[1]}_{-\frac{1}{2}} \ket*{\dot{A}}^{[1]}_1 G^{-B[2]}_{-\frac{1}{2}} \ket*{\dot{B}}^{[2]}_1 - G^{-B[1]}_{-\frac{1}{2}} \ket*{\dot{A}}^{[1]}_1 G^{-A[2]}_{-\frac{1}{2}} \ket*{\dot{B}}^{[2]}_1\right)\right]. 
\end{align}
There are three singleton global primaries corresponding to $3\dot{\chi}_1\phi_{0,2}^{(\rm{I})}$ in \eqref{eq:tower1'}:
\begin{align}
    \ket{1}^a&= \left(\sigma^a\right)_{\dot{A}\dot{B}} \left(J^{+(\rm{T})}_{-1} J^{-(\rm{T})}_0 \psi^{-\dot{A}(\rm{T})}_{-\frac{1}{2}} -J^{-(\rm{T})}_{-1} \psi^{+\dot{A}(\rm{T})}_{-\frac{1}{2}} +\epsilon_{AB} \alpha^{\dot{A}A(\rm{T})}_{-1} G^{-B(\rm{T})}_{-\frac{1}{2}}\right) (\ket{-}_1 \ket*{\dot{B}}_1)^{\ytableausetup{boxsize=1.2mm}\ydiagram[ ]{1,1}},
    \\
    \ket{2}^a&=\left(\sigma^a\right)_{\dot{A}\dot{B}} 
    \Bigl(
      2\psi^{-\dot{A}(\rm{T})}_{-\frac{3}{2}} 
      +2\epsilon_{\dot{C}\dot{D}} \psi^{-\dot{A}(\rm{T})}_{-\frac{1}{2}} \psi^{+\dot{C}(\rm{T})}_{-\frac{1}{2}} \psi^{-\dot{D}(\rm{T})}_{-\frac{1}{2}} 
      \notag    \\
    &\qquad\qquad\quad
      +3J^{+(\rm{T})}_{-1} J^{-(\rm{T})}_0 \psi^{-\dot{A}(\rm{T})}_{-\frac{1}{2}} 
      +4J^{3(\rm{T})}_{-1} \psi^{-\dot{A}(\rm{T})}_{-\frac{1}{2}}
    -5J^{-(\rm{T})}_{-1} \psi^{+\dot{A}(\rm{T})}_{-\frac{1}{2}} 
    \notag\\
    &\qquad\qquad\quad
    +\epsilon_{AB} G^{+A(\rm{T})}_{-\frac{1}{2}} G^{-B(\rm{T})}_{-\frac{1}{2}} \psi^{-\dot{A}(\rm{T})}_{-\frac{1}{2}} 
    -\epsilon_{AB} G^{-A(\rm{T})}_{-\frac{1}{2}} G^{-B(\rm{T})}_{-\frac{1}{2}} \psi^{+\dot{A}(\rm{T})}_{-\frac{1}{2}}
    \Bigr) 
    (\ket{-}_1 \ket*{\dot{B}}_1)^{\ytableausetup{boxsize=1.2mm}\ydiagram[ ]{1,1}},
    \\
    \ket{3}^a&=\left(\sigma^a\right)_{\dot{A}\dot{B}} \left(\epsilon_{\dot{C}\dot{D}} \psi^{-\dot{A}(\rm{T})}_{-\frac{1}{2}} \psi^{+\dot{C}(\rm{T})}_{-\frac{1}{2}} \psi^{-\dot{D}(\rm{T})}_{-\frac{1}{2}} -J^{+(\rm{T})}_{-1} J^{-(\rm{T})}_0 \psi^{-\dot{A}(\rm{T})}_{-\frac{1}{2}}\right.\notag\\
    &\qquad\qquad\qquad\qquad\qquad\left.+2J^{3(\rm{T})}_{-1} \psi^{-\dot{A}(\rm{T})}_{-\frac{1}{2}} +J^{-(\rm{T})}_0 \psi^{+\dot{A}(\rm{T})}_{-\frac{3}{2}}\right) (\ket{-}_1 \ket*{\dot{B}}_1)^{\ytableausetup{boxsize=1.2mm}\ydiagram[ ]{1,1}}.
\end{align}
Orthogonal linear combinations are
\begin{align}
    \big| s^{(\rm{I})}_{0,2} \big>^a &= \ket{1}^a,\\
    \big| s'^{(\rm{I})}_{0,2} \big>^a &= \ket{2}^a-\frac{17}{5}\ket{1}^a,\\
    \big| s''^{(\rm{I})}_{0,2} \big>^a &= \ket{3}^a+\frac{5}{8}\ket{1}^a-\frac{1}{8}\ket{2}^a,
\end{align}
which are used in \eqref{eq:0,2antiorthogonal}.

% \section{Lifting}

% inputs

\section{Characters and partition functions} \label{app:chars}

\def\fugoneL{{\eta}}
\def\fugtwoL{{\etad}}

Here we list partition functions and the characters for various algebras, defined by the NS-sector trace
\begin{align} \label{eq:Trace_def}
    {\rm Tr}\big[(-1)^{F}q^{L_0}y^{2J^3_0}\fugoneL^{2K^3_{1,0}}\fugtwoL^{2K^3_{2,0}}\big]
\end{align}
where $F$ is the fermion number, taken over the relevant representation space, except that we remove the overall factor of $(-1)^F$ so that the term of lowest conformal weight has a positive sign.
Note that we have included fugacities $\fugoneL$ and $\fugtwoL$ for $su(2)_1$ and $su(2)_2$, respectively.

\subsubsection*{Seed-theory partition function}

The left-moving (signed) partition function of the seed supersymmetric sigma model on $T^4$ in the NS sector is given by
% \begin{align} \label{eq:seed_z(q,y)}
%     z_{\mathrm{seed}}(q,y) \equiv (y+y^{-1}-\etad-\etad^{-1}) \prod_{r=1}^\infty 
%     {(1-yq^{r})^2(1-y^{-1}q^{r})^2\over (1-q^r)^4} 
%     = \sum_{r,\ell} c(r,\ell)\,q^r y^{\ell} \ ,
%\end{align}
\begin{align} \label{eq:seed_z(q,y)_NS}
    z_{\mathrm{seed}}^{\rm NS}(q,y,\eta,\etad) =\prod_{r=1}^\infty 
    {(1-y\etad q^{r-\half})(1-y\etad^{-1} q^{r-\half})(1-y^{-1}\etad q^{r-\half})(1-y^{-1}\etad^{-1} q^{r-\half})\over (1-\eta\etad q^r)(1-\eta \etad^{-1}q^r)(1-\eta^{-1}\etad q^r)(1-\eta^{-1}\etad^{-1}q^r)}.
\end{align}
The Ramond sector version is
\begin{align} \label{eq:seed_z(q,y)}
    z_{\mathrm{seed}}^{\rm R}(q,y,\eta,\etad) &= (y+y^{-1}-\etad-\etad^{-1}) \prod_{r=1}^\infty 
    {(1-y\etad q^{r})(1-y\etad^{-1} q^{r})(1-y^{-1}\etad q^{r})(1-y^{-1}\etad^{-1} q^{r})\over (1-\eta\etad q^r)(1-\eta \etad^{-1}q^r)(1-\eta^{-1}\etad q^r)(1-\eta^{-1}\etad^{-1}q^r)}
    \notag\\
    &\equiv \sum_{r,\ell} c(r,\ell;\eta,\etad)\,q^r y^{\ell} \ ,
\end{align}
where the expansion coefficients $c(r,\ell;\eta,\etad)$ in fact do not depend separately on $r$ and $\ell$ but only through one combination~\cite{Dijkgraaf:1996xw}: \textit{i.e.}\ $c(r,\ell;\eta,\etad)=c(4r-\ell^2;\eta,\etad)$.  Some first nonvanishing values of $c(x)=c(x;\eta,\etad)$ are $c(-1)=1$, $c(0)=-\dot\chi_\half$ and $  c(3)=1+\chi_\half\dot{\chi}_\half+\dot{\chi}_1$, where $\chi_j=\chi_j(\eta),\dot{\chi}_j=\chi_j(\etad)$ are defined in \eqref{eq:def_chi_j}.

\subsubsection*{$SU(2)$ character}

For $j=0,\half,1,\dots$, we define
\begin{align}
    \chi_j(z)=z^{2j}+z^{2(j-1)}+\cdots+z^{-2j}
        ={z^{2j+1}-z^{-2j-1}\over z-z^{-1}} \ .
        \label{eq:def_chi_j}
\end{align}
We often use the shorthand notation $\chi_j\equiv \chi_j(\fugoneL)$,  $\dot\chi_j\equiv\chi_j(\fugtwoL)$.  
We define $\chi_j(z)$ without an overall sign factor for any variables; for example, $\dot\chi_\half\equiv \chi_\half(\etad)=\etad+\etad^{-1}$.

\subsubsection*{$SU(1,1|2)$ character}

The character for a short global multiplet built on a chiral state of charges $h=j$ is
\begin{align} \label{eq.SU112_short_def}
\phi_{j}(q,y,\fugoneL)
 ={q^j\over 1-q}{
 y^{2j}[y-(\fugoneL+\fugoneL^{-1})q^{\half}+y^{-1}q]
 -y^{-2j}[y^{-1}-(\fugoneL+\fugoneL^{-1})q^{\half}+yq]
 \over  y-y^{-1}}.
\end{align}
The character for a long global multiplet built on a global primary of charges $(j,h)$, $j<h$, is
\begin{align} \label{eq.SU112_long_def}
 \phi_{j,h}(q,y,\fugoneL)
 &={q^h\over 1-q}
 {(y^{2j+1}-y^{-2j-1})\big[y^{-1}-(\fugoneL+\fugoneL^{-1})q^{\half}+qy\big]\big[y-(\fugoneL+\fugoneL^{-1})q^{\half}+qy^{-1}\big]\over y-y^{-1}}.
\end{align}
Our convention is that we do not have the overall factor $(-1)^{2j}$. 

In section~\ref{ssec:L_results} we will need to decompose products of short $SU(1,1|2)$ characters, as well as short characters with fugacities squared, back into short and long characters. These are given by
\begin{equation} \label{eq.SU112CharProd}
    \phi_{j_1}(q,y,\eta)\times \phi_{j_2}(q,y,\eta) = \phi_{j_1+j_2}(q,y,\eta) + \sum_{h\geq j_1+j_2} \sum_{j=|j_1-j_2|}^{j_1+j_2-1} \phi_{j,h}(q,y,\eta) \ ,
\end{equation}
where $h=j_1+j_2, j_1+j_2+1, j_1+j_2+2, \dots$ is integer if $j_1+j_2\in\mathbb{Z}$ and half-integer if $j_1+j_2\in\mathbb{Z}+\frac12$. Likewise
\begin{equation} \label{eq.SU112CharPow}
    \phi_{j}^{\ev{2}} \equiv \phi_{j}(q^2,y^2,\eta^2) = \phi_{2j}(q,y,\eta) + \sum_{h\geq 2j} \sum_{j'=0}^{2j-1} (-1)^{h+j'} \phi_{j',h}(q,y,\eta) \ .
\end{equation}

\subsubsection*{Contracted large $\cN=4$ character}

The characters for the contracted large $\cN=4$ algebra with central charge $c=6N$ were obtained in \cite{Petersen:1989pp, Petersen:1989zz}.  A generalized version with fugacities for $su(2)_1,su(2)_2$ in the NS sector is given by, for the short multiplet,
\begin{align}
&\Phibf_{j;N}(q,y,\fugoneL,\fugtwoL)
=
%(-1)^{2j}q^j
F(q,y,\fugoneL,\fugtwoL)
\notag\\
 &
 \times \sum_{n\in\bbZ}q^{N n^2 + (2j + 1) n}
\left[
\frac{ y^{2N n + 2j}}{(1-q^{n+\half}y \fugoneL )(1-q^{n+\half}y \fugoneL^{-1})}
-\frac{ y^{-2N n - 2(j + 1)}}{(1-q^{n+\half}y^{-1} \fugoneL)(1-q^{n+\half}y^{-1} \fugoneL^{-1})}\right],
\label{Phi(s)_N=4}
\end{align}
where $j=0,\frac12,1,\dots,{N\over 2}-{1\over 2}$, and for the long multiplet,
\begin{align}
\Phibf_{h,j;N}(q,y,\fugoneL, \fugtwoL)
=
%(-1)^{2j}
q^h F(q,y,\fugoneL, \fugtwoL)
 \times \sum_{n\in\bbZ}q^{N n^2 + (2j + 1) n}
 (y^{2N n + 2j}-y^{-2N n -2(j + 1 )}) \ ,
\label{Phi(l)_N=4}
\end{align}
where $j=0,\frac12,1,\dots,{N\over 2}-1$.
Here,
\begin{align}
F(q,y,\fugoneL,\fugtwoL)
&=\prod_{n=1}^\infty {
 (1-y \fugoneL q^{n-\half})
 (1-y^{-1} \fugoneL^{-1}q^{n-\half})
 (1-y^{-1} \fugoneL q^{n-\half})
 (1-y \fugoneL^{-1}q^{n-\half})
 \over (1-q^n)^2 (1-y^2 q^n)(1-y^{-2} q^{n-1})
 }
\notag\\
 &\quad\times\prod_{n=1}^\infty {
 (1-y \fugtwoL  q^{n-\frac12})
 (1-y^{-1} \fugtwoL^{-1}q^{n-\frac12})
 (1-y^{-1}\fugtwoL q^{n-\frac12})
 (1-y \fugtwoL^{-1}q^{n-\frac12})
\over 
 (1-\fugoneL \fugtwoL q^n)
 (1-\fugoneL^{-1} \fugtwoL^{-1} q^n)
 (1-\fugoneL^{-1} \fugtwoL q^n)
 (1-\fugoneL \fugtwoL^{-1} q^n)
 }\ ,
\end{align}
is a freely generated part by the action of $L_{-n},G^{\alpha A}_{-(n+\frac12)},J^a_{-n}$ (first line) and $\alpha^{\Ad A}_{-n},\psi^{\alpha \Ad}_{-(n-\frac12)}$ (second line).  In \eqref{Phi(s)_N=4} and \eqref{Phi(l)_N=4}, the infinite sum takes care of the singular vectors reached by the action of $G,J$. Again, our convention is that we do not have the overall factor $(-1)^{2j}$.

In the main text where we always assume $N=2$, we drop the subscript $N$ on $\Phibf_{j;N}$ and $\Phibf_{j,h;N}$.

%%%%%%%%%%%%%%%%%%%%%%%%%%%%%%%%%%%%%%%%%%%%%%%%%%%%%%%%%%%%%%%%%%%%%%%%%%%%%%
\bibliographystyle{utphys}
\bibliography{bib}
\end{document}